\documentclass[fleqn,usenatbib]{mnras}

\usepackage{newtxtext,newtxmath}
\usepackage[T1]{fontenc}

\DeclareRobustCommand{\VAN}[3]{#2}
\let\VANthebibliography\thebibliography
\def\thebibliography{\DeclareRobustCommand{\VAN}[3]{##3}\VANthebibliography}

\usepackage{graphicx}	% Including figure files
\usepackage{subcaption}	% 
\usepackage{amsmath}	% Advanced maths commands
\usepackage{siunitx}    % units
\usepackage{multirow}

\newcommand{\guvb}{$\Gamma_{\text{UVB}}$}
\newcommand{\lya}{Ly$\alpha$\ }
\newcommand{\snr}{\text{SNR}_{10}}

\DeclareSIUnit \h {h}
\DeclareSIUnit \parsec {pc}

\title[IGM thermal state from Lya auto-correlation]{Forecasting constraints on the high-$z$ IGM thermal state from the Lyman-$\alpha$ forest flux auto-correlation function
}

\author[Molly Wolfson et al.]{
Molly Wolfson$^{1}$\thanks{E-mail: mawolfson@ucsb.edu},
Joseph F. Hennawi$^{1,2}$,
Frederick B. Davies$^{3}$,
Zarija Luki\'{c}$^{4}$,
and Jose O{\~{n}}orbe$^{5}$
\\
$^{1}$Department of Physics, University of California, Santa Barbara, CA 93106, USA\\
$^{2}$Leiden Observatory, Leiden University, Niels Bohrweg 2, 2333 CA Leiden, Netherlands\\
$^{3}$Max-Planck-Institut f\"{u}r Astronomie, K\"{o}nigstuhl 17, 69117 Heidelberg, Germany\\
$^{4}$Lawrence Berkeley National Laboratory, 1 Cyclotron Rd, Berkeley, CA 94720, USA\\
$^{5}$Facultad de F\'{i}sicas, Universidad de Sevilla, Avda. Reina Mercedes s/n, Campus de Reina Mercedes, 41012 Sevilla, Spain
}

\date{Accepted XXX. Received YYY; in original form ZZZ}

\pubyear{2023}

\begin{document}
\label{firstpage}
\pagerange{\pageref{firstpage}--\pageref{lastpage}}
\maketitle

\begin{abstract}
% The abstract should be a single paragraph not more than 250 words (200 words for Letters).
The auto-correlation function of the Lyman-$\alpha$ (Ly$\alpha$) forest flux from high-z quasars can statistically probe all scales of the intergalactic medium (IGM) just after the epoch of reionization.
The thermal state of the IGM, which is determined by the physics of reionization, sets the amount of small-scale power seen in the \lya forest. 
To study the sensitivity of the auto-correlation function to the thermal state of the IGM, we compute the auto-correlation function from cosmological hydrodynamical simulations with semi-numerical models of the thermal state of the IGM.
We create mock data sets of 20 quasars to forecast constraints on $T_0$ and $\gamma$, which characterize a tight temperature-density relation in the IGM, at $5.4 \leq z \leq 6$.
At $z = 5.4$ we find that an ideal data set constrains $T_0$ to 29\% and $\gamma$ to 9\%.
In addition, we investigate four realistic reionization scenarios that combine temperature and ultra-violet background (UVB) fluctuations at $z = 5.8$.
We find that, when using mock data generated from a model that includes temperature and UVB fluctuations, we can rule out a model with no temperature or UVB fluctuations at $>1\sigma$ level 50.5\% of the time. 
\end{abstract}

\begin{keywords}
intergalactic medium -- dark ages, reionization, first stars -- quasars: absorption lines -- methods: statistical
\end{keywords}

%%%%%%%%%%%%%%%%%%%%%%%%%%%%%%%%%%%%%%%%%%%%%%%%%%

%%%%%%%%%%%%%%%%% BODY OF PAPER %%%%%%%%%%%%%%%%%%

\section{Introduction}

Understanding the epoch of reionization, the time period where the first luminous sources emitted ionizing that re-ionized the intergalactic medium (IGM), remains a major open problem about the early universe. 
The midpoint of reionization has been constrained from the cosmic microwave background to be $z_{\text{re}} = 7.7 \pm 0.7$ \citep{planck_2018}. 
Initial measurements of transmission in the Lyman-$\alpha$ (Ly$\alpha$) forest \citep{gunn_peterson_1965, lynds_1971} of high redshift quasars suggested that reionization was complete by $z \sim 6$ \citep{fan_2006, mcgreer_2011, mcgreer_2015}. 
Additional methods used to constrain reionization include observations of \lya emission from high redshift galaxies \citep[see e.g.][]{jung_2020, moreales_2021} and large \lya absorption troughs \citep[see e.g.][]{becker_2018, kashino_2020}.
Measurements of the \lya forest optical depths scatter on levels that suggest reionization is not actually complete until $z < 6$ \citep{fan_2006, becker_2015, bosman_2018, eilers_2018, yang_2020, bosman_2021_data}.
An alternative, indirect method to constrain reionization is by looking at the thermal history of the IGM at $z > 5$ \citep{boera_2019, walther_2019, gaikwad_2020}.

During reionization, ionization fronts propagate through the IGM and impulsively heat the reionized gas in the IGM to $\sim 10^4$ K \citep{mcquinn_2012, davies_2016, daloisio_2019}.
The details of the driving sources, the timing, and duration of reionization will determine the precise amount of heat injected. 
After reionization, the IGM expands and cools through the adiabatic expansion of the universe and inverse Compton scattering off CMB photons. 
The combination of these physical processes will allow the IGM gas to relax into a state described by a tight power-law relation between the temperature and density: 
\begin{equation}
    T = T_0 \Delta^{\gamma - 1}.
    \label{eq:temperature-density relation}
\end{equation}
Where $\Delta = \rho / \bar{\rho}$ is the overdensity, $\bar{\rho}$ is the mean density of the Universe, $T_0$ is the temperature at mean density, and $\gamma$ is the slope of the relationship \citep{hui_gnedin_1997, puchwein_2015, mcquinn_upton_2016}. 
The low-density IGM has long cooling times, so the thermal memory of reionization will persist for hundreds of Myr.
This means that thermal state of the IGM at the end and after reionization, $z \sim 5 - 6$, can provide key insights into reionization \citep{miralda_escude_1994, hui_gnedin_1997, haehnelt_1998, theuns_2002_wavelet, hui_haiman_2003, lidz_2014, onorbe_2017, onorbe_2017_planck}.

The Ly$\alpha$ optical depth, $\tau_{\rm Ly\alpha}$ is related to the temperature via 
\begin{equation}
    \tau_{\rm Ly\alpha} = n_{\rm HI} \sigma_{{\rm Ly}\alpha} \propto  T^{-0.7} / \Gamma_{\rm HI},
    \label{eq:optical_depth_temp_uvb}
\end{equation}
see \citet{Rauch_1998}. 
Thus, several statistics have been used to measure the thermal state of the IGM from the \lya forest, including 
the flux probability density \citep{becker_2007, bolton_2008, viel_2009, calura_2012, lee_2015}, the curvature \citep{becker_2011, boera_2014, gaikwad_2020}, 
the Doppler parameter distribution \citep{schaye_1999, schaye_2000, ricotti_2000, bryan_2000, mcdonald_2001, rudie_2012, bolton_2010, bolton_2012, bolton_2014, rorai_2018, gaikwad_2020}, 
the joint distribution of the Doppler parameters with the Hydrogen Column Density \citep{hiss_2018}, 
and wavelets \citep{lidz_2010, garzilli_2012, gaikwad_2020}. 
One of the most commonly used statistics used to measure the structure of the Ly$\alpha$ forest is the 1D flux power spectrum, $P_{F}(k)$ \citep{theuns_2000, zaldarriaga_2001, yeche_2017, walther_2017, boera_2019, gaikwad_2020}. 

The thermal state of the IGM will affect the \lya forest via Doppler broadening due to thermal motions and Jeans (pressure) smoothing of the underlying baryon distribution.
The physics of Jeans smoothing are as follows. 
The rate at which pressure forces erase gravitational fluctuations is set by the local sound speed.
At IGM densities the pressure scale sound crossing time is approximately the Hubble time so the pressure smoothing scale provides an integrated record of the thermal history of the IGM \citep{gnedin_hui_1998, kulkarni_2015, nasir_2016, onorbe_2017, onorbe_2017_planck, rorai_2017}.
Both Doppler broadening and Jeans smoothing reduce the small-scale structure of the \lya forest.
These reductions in small-scale structure of the \lya forest lead to a cut-off in $P_{F}(k)$ at high-$k$.

An alternative to the power spectrum is the \lya forest flux auto-correlation function, which is the Fourier transform of the power spectrum. 
In this work we will explore the ability of the auto-correlation function to constrain the thermal state of the IGM at $z > 5$. 
The auto-correlation function of the \lya forest has two statistical properties that make it easier to work with than the power spectrum.
First is that uncorrelated noise (which is the expectation for astronomical spectrograph noise) will not impact non-zero lags of the auto-correlation function, as it will average to zero. 
This allows the auto-correlation function to automatically remove noise, while uncorrelated noise is a constant positive value at all scales for the power spectrum. 
Thus the unknown noise level must be calculated and subtracted from power spectrum measurements which will add additional uncertainty to the final measurement. 
Additionally, observational quasar spectra often have regions that need to be removed (e.g. for metal lines). 
Masking out these and other regions introduces a complicated window function to the power spectrum that must be corrected for (see e.g. \citet{walther_2019}) and will again increase the uncertainty in the measurement. 
The auto-correlation function does not require a similar correction since masking will only change the number of pixel pairs used at a given velocity lag. 

Many previous studies have measured the \lya forest flux auto-correlation function at lower redshifts for a wide range of applications \citep{mcdonald_2000, rollinde_2003, becker_2004, dodorico_2006}. 
In addition, the first measurement of the \lya forest flux auto-correlation function at $z > 5$ is presented in \citet{wolfson_2023_xqr30} for moderate resolution quasar spectra. 

In this work we will investigate constraints on $T_0$ and $\gamma$ that can be achieved from measurements of the \lya forest flux auto-correlation function. 
We will do this by creating mock observational measurements of the auto-correlation function and comparing to model values of the auto-correlation function determined via semi-numerical methods applied to hydrodynamical simulations. 
By applying Bayesian statistics to this setup we will get mock posterior distributions for $T_0$ and $\gamma$. 

Beyond a thermal state that follows a tight power law described by $T_0$ and $\gamma$, reionization can lead to significant fluctuations in the temperature of the IGM \citep{daloisio_2015, davies_2018_thermal}.
At the same time, the existence of significant neutral hydrogen in the IGM can result in large spatial fluctuations in the  ultraviolet background (UVB) \citep{davies_furlanetto_2016, gnedin_2017,daloisio_2018}. 
Fluctuations in the UVB arise during reionization because the ionizing photons produced will be absorbed by the remaining neutral hydrogen short distances from their initial sources. 
This distance is characterized by the mean free path of ionizing photons, $\lambda_{\text{mfp}}$ \citep{mesinger_furlanetto_2009}.
Various previous studies have investigated the effect of large scale variations in the UVB on the auto-correlation function and power spectrum of the \lya forest \citep{zuo_1992_a, zuo_1992_b, croft_2004, meiksin_2004, mcdonald_2005, gontcho_2014, pontzen_2014_a, pontzen_2014_b, daloisio_2018, meiksin_2019, onorbe_2019}. 
In particular, \citet{wolfson_2022} showed that the positive fluctuations in the UVB that accompany small $\lambda_{\text{mfp}}$ values boost the flux of the \lya forest on small scales, which can be detected in the auto-correlation function.

We will use an additional hydrodynamical simulation that models fluctuations in both the temperature and the UVB to determine the effect on the \lya forest flux auto-correlation function. 
In addition to looking at the qualitative differences between these models, we will quantify the likelihood ratio between different models for mock data sets. 
This provides a quantitative way to discuss constraints on a discrete set of models.

The structure of this paper is as follows. 
We discuss our grid of simulations that vary $T_0$ and $\gamma$ in Section \ref{section: sims}. 
The auto-correlation function and our other statistical methods to constrain these parameters are described in Section \ref{section: methods} with our results being discussed in Section \ref{section: thermal results} 
We discuss our second set of simulations for models of the IGM with temperature and UVB fluctuations in Section \ref{section: inhomo reion} and use the auto-correlation function to quantitatively distinguish between these models in Section \ref{sec:ruling out}. 
Finally, we summarize in Section \ref{section: conclusion}.

\section{Simulation Data} \label{section: sims}

\subsection{Simulation box} \label{subsec: sims}

In this work we use a simulation box of size $L_{\text{box}} = 100$ comoving Mpc (cMpc) h$^{-1}$ run with \texttt{Nyx} code \citep{almgren_2013}. 
\texttt{Nyx} is a hydrodynamical simulation code that was designed for simulating the Ly$\alpha$ forest with updated physical rates from \citet{lukic_2015}.
The simulation has with $4096^3$ dark matter particles and $4096^3$ baryon grid cells. 
It is reionized by a \citet{haardt_madau_2012} uniform UVB that is switched on at $z \sim 15$. 
We have two snapshots of this simulation at $z = 5.5$ and $z = 6.0$. 
We consider models at seven redshifts: $5.4 \leq z \leq 6.0$ with $\Delta z = 0.1$.
In order to consider the redshifts for which we do not have a simulation output, we select the nearest snapshot and use the desired redshift when calculating the proper size of the box and the mean density. 
This means we use the density fluctuations and velocities directly from the nearest \texttt{Nyx} simulation output. 
As a test, we used the $z=6.0$ simulation snapshot to generate skewers at $z = 5.7$ and found no significant change in our finally results, thus the nearest grid point interpolation between snapshot redshifts is sufficient.

\begin{figure}
	\includegraphics[width=\columnwidth]{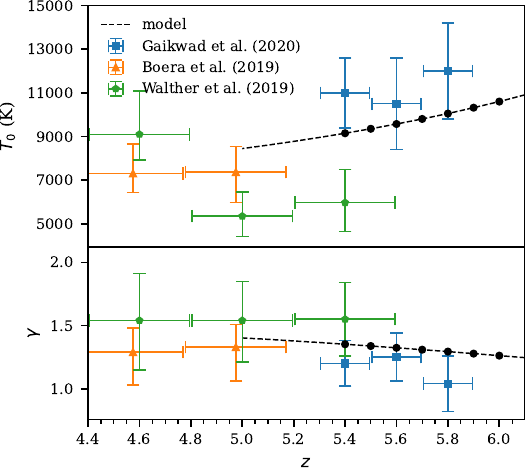}
    \caption{
        The blue squares and orange and green triangles show previous measurements of $T_0$ and $\gamma$ at high $z$ from \citet{gaikwad_2020}, \citet{boera_2019}, and \citet{walther_2019} respectively. 
        % The solid line shows the ``late reionization" model from \citet{onorbe_2017_model} while 
        The dashed line shows the results for a thermal evolution model calculated with methods similar to \citet{sanderbeck_2016} and \citet{davies_2018_thermal}
        This model has $z_{\text{reion}} = 7.7$, $\Delta T = 20,000$ K, and $\alpha_{\text{UVB}} = 1.5$. 
        We use this model as our ``true" redshift evolution for $T_0$ and $\gamma$ in this work. 
        The chosen models are shown as black circles. 
    }
    \label{fig:temp_gamma_model_evolution}
\end{figure}

We generate grids of thermal models through a semi-numerical method to set the temperature along the sightlines.
For each value of $T_0$ and $\gamma$, we set the temperature of each cell following Equation \eqref{eq:temperature-density relation} for all densities with no cutoff.
Our method does not take into account the full evolution of the thermal state of the IGM, only the instantaneous temperature. 
This simple model is sufficient to achieve the aim of this paper, which is to see if the auto-correlation function is sensitive to the thermal state. 
To make our grid we use 15 values of $T_0$ and 9 values of $\gamma$ resulting in 135 different combinations of these parameters at each $z$. 
The values of $T_0$ and $\gamma$ in our grid of thermal models were chosen based on the current models and available data, as shown in Figure \ref{fig:temp_gamma_model_evolution}. 
We generate a model for the evolution of the thermal state of the IGM by a method similar to \citet{sanderbeck_2016} with $z_{\text{reion}} = 7.7$, $\Delta T = 20,000$ K, and $\alpha_{\text{UVB}} = 1.5$. 
For more information on the calculation of the temperature field see \citet{davies_2018_thermal}.
We select central ``true'' $T_{0}$ and $\gamma$ values at each redshift from this model, which are shown as black points in Figure \ref{fig:temp_gamma_model_evolution} and listed in Table \ref{tab:central vals}.
At all $z$, we use the errors on the measurements reported in \citet{gaikwad_2020} at $z = 5.8$ ($\Delta T_0 = \SI{2200}{\kelvin}$ and $\Delta \gamma = 0.22$) and modeled from $T_{0} - 4\Delta T_0$ to $T_{0} + 4\Delta T_0$ and $\gamma - 4\Delta\gamma$ to $\gamma + 4\Delta\gamma$ in linear bins.

\begin{table}
    \centering
    \caption{
        This table lists the central `true" values of the redshift-dependent thermal state models used in this work.
        % include the range somehow?
        The last column states the central ``true" value of $\langle F \rangle$ modeled in this work, which are the measurements from \citet{bosman_2021_data}. 
    }
	\label{tab:central vals}
    \begin{tabular}{|c|c|c|c|}
        \hline $z$ & $T_0$ (K) & $\gamma$ & $\langle F \rangle$ \\ \hline
        5.4 & 9,149     & 1.352    & 0.0801   \\
        5.5 & 9,354     & 1.338    & 0.0591   \\
        5.6 & 9,572     & 1.324    & 0.0447   \\
        5.7 & 9,804     & 1.309    & 0.0256   \\
        5.8 & 10,050    & 1.294    & 0.0172   \\
        5.9 & 10,320    & 1.278    & 0.0114   \\
        6.0 & 10,600    & 1.262    & 0.0089   \\ \hline
    \end{tabular}
\end{table}

Our simulations don't predict the overall average of the UVB, $\overline{\Gamma_{\text{UVB}}}$, because this value originates from complicated galaxy physics that are not included in the simulations. 
In addition our method of post-processing different thermal states would affect the resulting $\overline{\Gamma_{\text{UVB}}}$.
Instead we choose to model a variety of potential $\overline{\Gamma_{\text{UVB}}}$ values through the mean transmitted flux, $\langle F \rangle$.
This is done be re-scaling the optical depths along the skewer, $\tau$, such that $\langle e^{-\tau} \rangle = \langle F \rangle$ when averaging the transmitted flux over all skewers.
These $\langle F \rangle$ model values are centered on the values presented in \citet{bosman_2021_data} for each redshift bin. 
We chose a range of models spanning $4\Delta \langle F \rangle$ where the $\Delta \langle F \rangle$ is the redshift dependent value reported in \citet{bosman_2021_data}. 
These choices of $\langle F \rangle$ are listed in the last column of Table \ref{tab:central vals}.

\begin{figure*}
	\includegraphics[width=2\columnwidth]{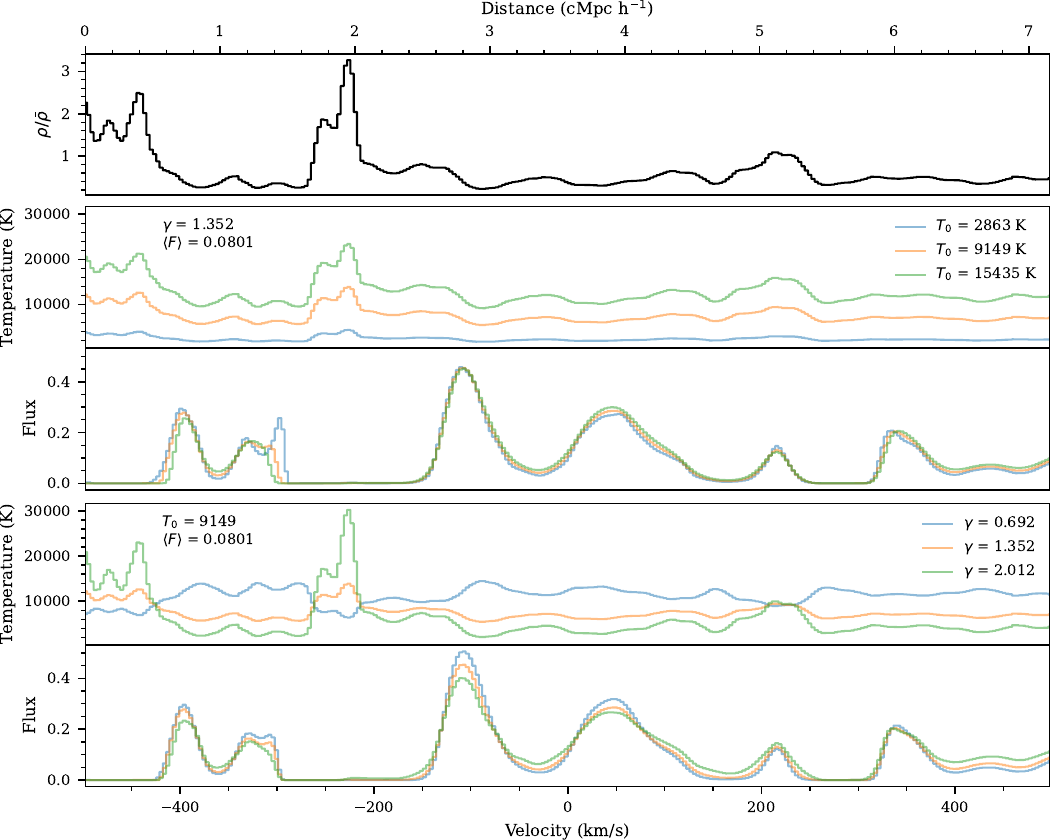}
    \caption{
        The top panel shows the density along a small section of one skewer in black. 
        This skewer is taken at $z = 5.4$.
        There are then two pairs of panels each depicting the temperature (top) and flux (bottom) along this skewer. 
        The first pair varies $T_0$ with constant $\gamma = 1.352$ and $\langle F \rangle =0.0801$. 
        Shifting $T_0$ causes a corresponding shift in the temperature values along the skewer. 
        Hotter temperatures (orange and green) smooths the flux, as seen clearly in the loss of a second transmission spike at $z \sim \SI{-300}{\kilo\meter\per\second}$. 
        The second pair varies $\gamma$ with constant $T_0 = \SI{9148}{\kelvin}$ and $\langle F \rangle = 0.0801$.
        When $\gamma > 1$ (orange and green) the temperature is directly proportional to the density fluctuations while $\gamma < 1$ (blue) causes the temperature to be inversely proportional to the density fluctuations. 
        When temperature is inversely proportional to density, lower densities have higher temperatures.
        Low densities and higher temperatures will locally increase the flux so the $\gamma < 1$ (blue) model will lead to transmission spikes with the greatest flux, as seen at $v \sim \SI{-100}{\kilo\meter\per\second}$.
        }
    \label{fig:flux_two_paenl}
\end{figure*}

We draw 1000 skewers from the simulation box.
One example skewer at $z = 5.4$ for different $T_0$ and $\gamma$ models is shown in Figure \ref{fig:flux_two_paenl}. 
The top panel shows the density of this skewer for all models in black.
There are then two pairs of panels each depicting the temperature (top) and flux (bottom) along this skewer. 
The second and third panels vary $T_0$ with constant $\gamma = 1.352$ and $\langle F \rangle = 0.0801$. 
The coldest model, $T_0 = \SI{2863}{\kelvin}$ (blue), has some of the sharpest features. 
This is seen at $v \sim \SI{-300}{\kilo\meter\per\second}$ where the low $T_0$ (blue) model has a secondary sharp peak in the flux. 
In comparison the hottest model, $T_0 = \SI{15435}{\kelvin}$ (green), has one wider transmission spike. 
In addition, increasing $T_0$ decreases $\tau_{\text{Ly}\alpha}$ as described in Equation \eqref{eq:optical_depth_temp_uvb}, which in turn increases the transmitted flux. 
For this reason we get the greatest transmission from the $T_0 = 15435$ K (green) model, seen in the transmission spike at $v = \SI{50}{\kilo\meter\per\second}$.
With fixed $\langle F \rangle$ this leads to greater variation in the flux for higher $T_0$ models.

The fourth and fifth panels vary $\gamma$ with constant $T_0 = 9149$ K and $\langle F \rangle = 0.0801$. 
When $\gamma > 1$ (orange and green) the temperature is directly proportional to the density fluctuations while $\gamma < 1$ (blue) causes the temperature to be inversely proportional to the density fluctuations. 
When temperature is inversely proportional to density, lower densities have higher temperatures.
Low densities and higher temperatures will locally increase the flux so the $\gamma < 1$ (blue) model will lead to transmission spikes with the greatest flux, as seen at $v \sim \SI{-100}{\kilo\meter\per\second}$.

\subsection{Forward Modeling} \label{section: forward modeling}

In order to mimic realistic observational data from echelle spectrographs, (e.g. from Keck/HIRES, VLT/UVES, and Magellan/MIKE) we forward model our ideal simulation skewers to have imperfect resolution and flux levels. 
We consider a resolution of $R = 30000$ and a signal to noise ratio per \SI{10}{\kilo\meter\per\second} pixel ($\snr$) of $\snr = 30$ at all redshifts. 

We model this resolution by smoothing the flux by a Gaussian filter with $\text{FWHM} = \SI{10}{\kilo\meter\per\second}$. 
After smoothing we re-sampled the new flux such that the new pixel size was $\Delta v = \SI{2.5}{\kilo\meter\per\second}$.
With this pixel scale, $\snr = 30$ corresponds to a signal to noise ratio of the pixel size ($\text{SNR}_{\Delta v}$) of 15. 
For simplicity, we add flux-independent noise in the following way.
We generate one 1000 skewer $\times$ skewer length realization of random noise drawn from a Gaussian with $\sigma_N = 1 / \text{SNR}_{\Delta v}$ and add this noise realization to every model at every redshift.
Using the same noise realization over the different models prevents stochasticity from different realizations of the noise from adding additional variations between the models. 
Thus the noise modeling will not unduly, corrupt the parameter inference.

As discussed in Section \ref{subsec: sims} simulation skewers are 100 cMpc h$^{-1}$ long, much longer than the $\Delta z = 0.1$ redshift bins we have chosen to analyze. 
Therefore, we split these skewers into two regions of length $\Delta z = 0.1$ and treating these two regions as independent, resulting in a total of 2000 skewers. 
Note that $\Delta z = 0.1$ corresponds to $33$ cMpc h$^{-1}$ at $z = 5.4$ and $29$ cMpc h$^{-1}$ at $z = 6.0$.

\begin{figure}
	\includegraphics[width=\columnwidth]{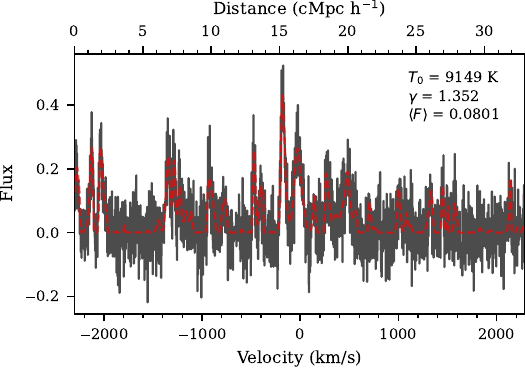}
    \caption{
        This shows a skewer at $z = 5.4$ with $T_0 = \SI{9149}{\kelvin}$, $\gamma = 1.352$, and $\langle F \rangle = 0.0801$ (our assumed ``true" parameters at this redshift).
        This skewer, as is true for all skewers, is forward modeled with $R = 30000$ and $\snr = 30$. 
        The initial flux is plotted as the red dashed line while the forward modeled flux is plotted as the black histogram. 
    }
    \label{fig:flux_noise}
\end{figure}

The initial and forward-modeled flux for one $z = 5.4$ skewer is shown in Figure \ref{fig:flux_noise}. 
This skewer has $T_0 = \SI{9149}{\kelvin}$, $\gamma = 1.352$, and $\langle F \rangle = 0.0801$ (our assumed ``true" parameters at this redshift).
The forward modeled skewer, as is always true, uses $R = 30000$ and $\snr = 30$. 
The initial flux is plotted as the red dashed line while the forward modeled flux is plotted as the black histogram. 

We assume a fiducial data set size of 20 quasar spectra that probe a redshift interval of $\Delta z = 0.1$ per quasar for a total pathlength of $\Delta z = 2.0$ at all redshifts. 
This is a reasonable number of high-$z$, high resolution quasar observations to consider for a future measurement.

\section{Methods} \label{section: methods}

\subsection{Auto-correlation} \label{sec:autocorr}

The auto-correlation function of the flux, $\xi_F (\Delta v)$, is defined as
\begin{equation}
    \xi_F (\Delta v) = \langle F(v) F(v + \Delta v) \rangle
    \label{eq:autocorr}
\end{equation}
where $F(v)$ is the flux of the \lya forest and the average is performed over all pairs of pixels at the same velocity lag, $\Delta v$. 
Conventionally, the flux contrast field, $\delta_F = (F - \langle F \rangle) / \langle F \rangle$, is used when measuring the power spectrum of the \lya forest. 
Here, we chose to use the flux since $\langle F \rangle$ is small and has large uncertainties at high-$z$ where we are most interested in this measurement.
Using the flux thus prevents us from dividing by a small number which comes from an independent measurement and could potentially blow up the value of the flux contrast. 
The auto-correlation function of the flux contrast can be written as 
\begin{equation}
    \xi_{\delta_f}(\Delta v) = \frac{\xi_F(\Delta v) - \langle F \rangle^2}{\langle F \rangle^2}.
    \label{eq:autocorr_deltaf}
\end{equation}
$\xi_{\delta_f}$ can be computed via the Fourier transform of the dimensionless power spectrum of the \lya forest flux contrast, $\Delta^2_{\delta_F}(k) = kP_{\delta_f}(k)/\pi$. 
In one dimension this can be written:
\begin{equation}
    \xi_{\delta_f}(\Delta v) = \int_0^{\infty} \Delta^2_{\delta_f}(k)\cos(k \Delta v) d \ln k
    \label{eq:autocorr_int_power}
\end{equation}
The dimensionless power, $\Delta^2_{\delta_f}(k)$, is a smoothly rising function that has a sharp cutoff set by the thermal state of the IGM. 
Higher temperature values lead to sharper cutoffs as the power at small scales in the \lya forest is removed. 
Equation \eqref{eq:autocorr_int_power} can be particularly useful when building intuition for the trends seen in the auto-correlation function with changing $T_0$ and $\gamma$, which we will discuss later in this section.

We compute the auto-correlation function with the following consideration for the velocity bins. 
We set the left edge of the smallest bin to be the resolution length, $\SI{10}{\kilo\meter\per\second}$, and continue with linear bin sizes with a width of the resolution length, $\SI{10}{\kilo\meter\per\second}$, up to $\SI{300}{\kilo\meter\per\second}$. 
Then we switch to logarithmic bin widths where $\log(\Delta v) = 0.029$ out to a maximal distance of $\SI{2700}{\kilo\meter\per\second}$. 
This results in 59 velocity bins considered where the first 28 have linear spacing. 
The center of our smallest bin is $\SI{15}{\kilo\meter\per\second}$ and the center of our largest bin is $\SI{2295}{\kilo\meter\per\second}$.
This largest bin corresponds to $\sim 16.5$ cMpc h$^{-1}$ at $z = 5.4$. 
We chose to use linear bins on the smallest scales because this is where the thermal state has the greatest effect on the \lya forest flux. 
At larger scales we switch to logarithmic binning as this is only sensitive to $\langle F \rangle$ and not the thermal parameters.
The main aim of this work is to constrain the thermal parameters so having fine binning at large scales is not as important. 
To check this we compared out results at $z = 5.4$ to those when using linear bins at all scales and found no significant change to the constraints on the parameters.
However, using linear bins at all scales results in 268 total bins, which significantly slowed down our computations.  
Therefore we used the linear-logarithmic bins at all $z$ throughout the rest of this work. 

The model value of the auto-correlation function was determined by taking the average of the auto-correlation function over all 2000 forward-modeled skewers.
Each mock data set of the auto-correlation was calculated by taking an average over 20 random skewers (representing 20 quasar sightlines) from the initial 2000 forward-modeled skewers.
The value of the auto-correlation function at the smallest velocity lags is affected by the finite resolution. 
This effect is left in both the models and the mock data.

\begin{figure*}
	\includegraphics[width=2\columnwidth]{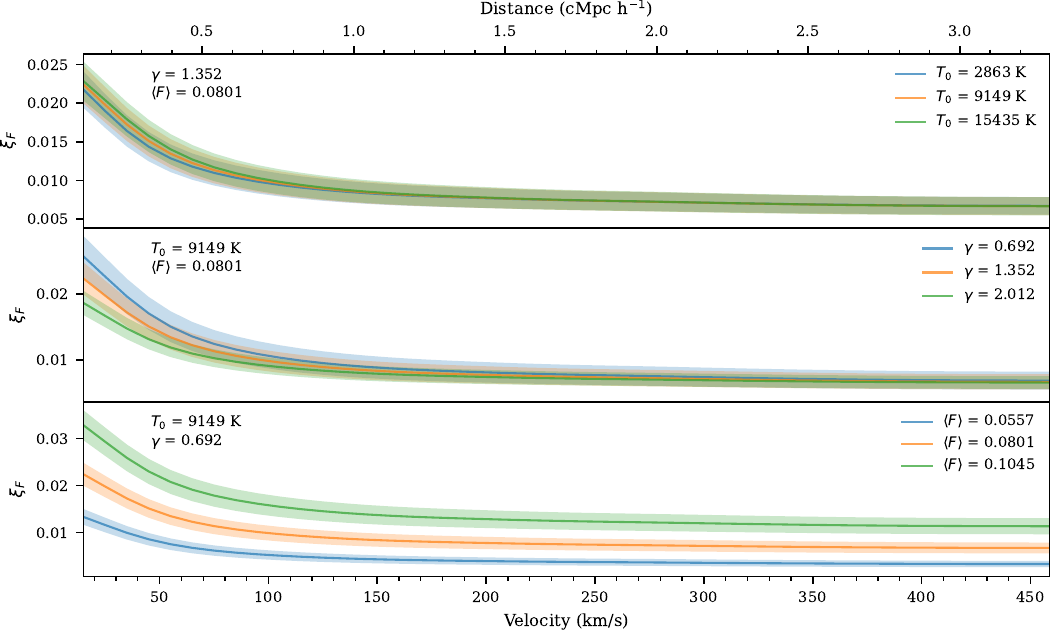}
    \caption{
        This figure demonstrates the effects of varying the parameters on the auto-correlation function from the simulations at $z = 5.4$. 
        The three panels varies one parameter while keeping the others constant from $T_0$, $\gamma$, and $\langle F \rangle$, respectively.
        The solid lines show the model values and the shaded regions show errors estimated by the diagonals of the covariance matrices. 
        $T_0$ (top panel) and $\gamma$ (middle panel) affect the auto-correlation function on small scales.  
        $\langle F \rangle$ (bottom panel) affects the auto-correlation function on all scales.
    }
    \label{fig:correlation_three_panels}
\end{figure*}

We show the correlation functions calculated for different thermal state parameters in Figure \ref{fig:correlation_three_panels} at $z = 5.4$. 
The solid lines show the mean values while the shaded regions represent the errors estimated from the diagonal of the covariance matrices. 
We discuss the computation of these covariance matrices later in this section. 

The top panel shows models that vary $T_0$ with constant $\gamma$ and $\langle F \rangle$.
Varying $T_0$ results in small changes for the smallest velocity lags, where the second bin centered on $\SI{25}{\kilo\meter\per\second}$ has the largest percent change in the models. 
The middle panel has models that vary $\gamma$ with constant $T_0$ and $\langle F \rangle$ where the effect of changing $\gamma$ is strongest on small scales.  
The bottom panel has models that vary $\langle F \rangle$ with constant $T_0$ and $\gamma$.
$\langle F \rangle$ sets the amplitude of the auto-correlation function at all velocity lags.
Here the differences between models are linear where larger $\langle F \rangle$ leads to larger auto-correlation values. 
This scaling is roughly $\propto \langle F \rangle^2$, which follows from the definition of the auto-correlation function. 

For the thermal models, larger $T_0$ and smaller $\gamma$ lead to larger correlation function values on small scales. 
Though these models do not seem to show large differences by eye, we will investigate what statistically rigorous measurements could look like in Section \ref{section: thermal results}. 

To build intuition for the behavior of the auto-correlation function with the thermal parameters we refer to Equation \eqref{eq:autocorr_int_power}. 
In Appendix \ref{appendix:power_cos} we show the integrand from this equation for $\Delta v = \SI{15}{\kilo\meter\per\second}$.
As mentioned above $\Delta^2_{\delta_f}(k)$ has a sharp thermal cutoff which would naively lead to the belief that the auto-correlation function could show lower values at small-scales for hotter thermal states. 
However, as seen in Figure \ref{fig:correlation_three_panels}, greater $T_0$ values have greater values of the auto-correlation function at small scales. 
This behavior is explained by the variation for the flux in these models, as described earilier in the section, but also from the behavior of $\Delta^2_{\delta_f}(k)$ at small $k$ values with a linear y-scale, as seen in Figure \ref{fig:app_power_cos}.

We compute the covariance matrices for the models from mock draws of the data:
\begin{equation}
    \Sigma(T_0, \gamma,\langle F \rangle) = \frac{1}{N_{\text{mocks}}} \sum_{i=1}^{N_{\text{mocks}}}(\boldsymbol{\xi}_i - \boldsymbol{\xi_\text{model}})(\boldsymbol{\xi}_i - \boldsymbol{\xi_\text{model}})^{\text{T}}
    \label{eq:covariance}
\end{equation}
where $\boldsymbol{\xi}_i =  \boldsymbol{\xi}_i(T_0, \gamma,\langle F \rangle)$ is the $i$-th mock auto-correlation function, $\boldsymbol{\xi_\text{model}} = \boldsymbol{\xi_\text{model}}(T_0, \gamma,\langle F \rangle)$ is the model value of the auto-correlation function, and $N_{\text{mocks}}$ is the number of forward-modeled mock data sets used.
We use $N_{\text{mocks}} = 500000$ for all models and redshifts in this work, see Appendix \ref{appendix:converge} for a discussion on the convergence of the covariance matrix.
Note that $ \boldsymbol{\xi}_i(T_0, \gamma,\langle F \rangle)$ and $\Sigma(T_0, \gamma,\langle F \rangle)$ are computed at each point on the grid of $T_0$, $\gamma$, and $\langle F \rangle$, resulting in 1215 separate computations.

\begin{figure}
	\includegraphics[width=\columnwidth]{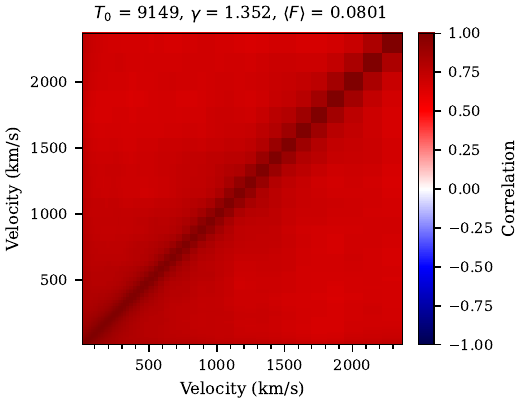}
    \caption{
        This figure shows the correlation matrix calculated with equation \eqref{eq:correlation} for the model at $z = 5.4$ with $T_0 = \SI{9149}{\kelvin}$, $\gamma = 1.352$, $\langle F \rangle = 0.0801$.
        The color bar is fixed to span from -1 to 1, which is all possible values of the correlation matrix. 
        This illustrates that all bins in the auto-correlation function are highly correlated with each other. 
    }
    \label{fig:covar_correlation}
\end{figure}

To visualize the covariance matrix, we define the correlation matrix, $C$, which expresses the covariances between $j$th and $k$th bins in units of the the diagonal elements of the covariance matrix. 
This is done for the $j$th, $k$th element by
\begin{equation}
        C_{jk} = \frac{\Sigma_{jk}}{\sqrt{\Sigma_{jj}\Sigma_{kk}}}.
        \label{eq:correlation}
\end{equation}
One example correlation matrix is shown in Figure \ref{fig:covar_correlation} for $z = 5.4$ with $T_0 = \SI{9149}{\kelvin}$, $\gamma = 1.352$, $\langle F \rangle = 0.0801$.
All bins of the auto-correlation function are very-highly correlated which is due to the fact that each pixel in the \lya forest contribute to multiple (in fact almost all) bins in the auto-correlation function.

\subsection{Parameter Estimation} \label{section: mcmc}

To quantitatively constrain the parameters we modeled ($T_0$, $\gamma$, and $\langle F \rangle$), we use Bayesian inference with a multivariate Gaussian likelihood and a flat prior over the parameters.
This likelihood, $\mathcal{L} = p(\boldsymbol{\xi} | T_0, \gamma, \langle F \rangle)$, has the form:
\begin{equation}
        \mathcal{L} = \frac{1}{\sqrt{\det(\Sigma) (2 \pi)^{n}}} \exp \left( -\frac{1}{2} (\boldsymbol{\xi} - \boldsymbol{\xi_\text{model}})^{\text{T}} \Sigma^{-1} (\boldsymbol{\xi} - \boldsymbol{\xi_\text{model}}) \right) 
        \label{eq:gauss_like}
\end{equation}
where $\boldsymbol{\xi}$ is the auto-correlation function from our mock data, 
$\boldsymbol{\xi_\text{model}}=\boldsymbol{\xi_\text{model}}(T_0, \gamma,\langle F \rangle)$ is the model value of the auto-correlation function,
$\Sigma = \Sigma(T_0, \gamma,\langle F \rangle)$ is the model dependent covariance matrix estimated by Equation \eqref{eq:covariance}, 
and $n=59$ is the number of points in the auto-correlation function. 
We discuss the assumption of using a multivariate Gaussian likelihood in Appendix \ref{appendix:multi gaussian}. 
This discussion shows that our mock data does not exactly follow a Guassian distribution.
This discrepancy may affect our parameter inference, we investigate the consequences of this assumption in a later section. 

Our models are defined by three ($T_0$, $\gamma$, and $\langle F \rangle$) parameters. 
We compute the posteriors on these parameters using Markov Chain Monte Carlo (MCMC) with the \texttt{EMCEE} \citep{forman_mackey_2013} package.
We linearly interpolate the model values and covariance matrix elements onto a finer 3D grid of $T_0$, $\gamma$, and $\langle F \rangle$ then use the nearest model during the MCMC. 
This fine grid has 29 values of $T_0$, 33 values of $\gamma$, and 41 values of $\langle F \rangle$ which corresponds to adding 1, 3, and 4 points between the existing grid points respectively.
Our MCMC was run with 16 walkers taking 3500 steps each and skipping the first 500 steps of each walker as a burn-in.

\begin{figure*}
    \centering
    
    \begin{subfigure}[t]{\textwidth}
    \centering
        \includegraphics[width=\linewidth]{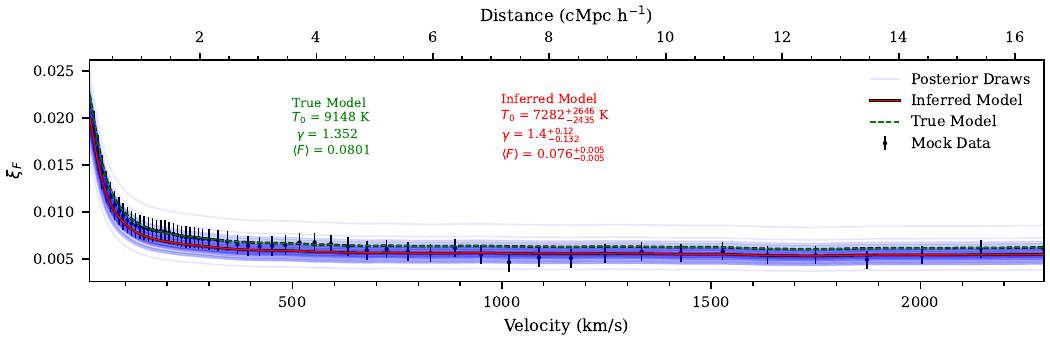} 
    \end{subfigure}
    
    \vspace{3mm}

    \begin{subfigure}[t]{0.49\textwidth}
        \centering
        \includegraphics[width=\linewidth]{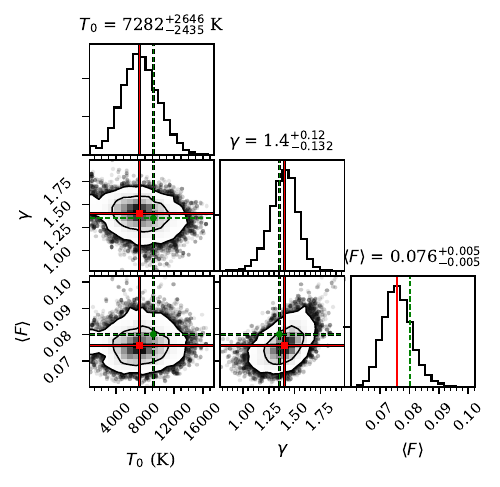} 
    \end{subfigure}
    \hfill
    \begin{subfigure}[t]{0.49\textwidth}
        \centering
        \includegraphics[width=\linewidth]{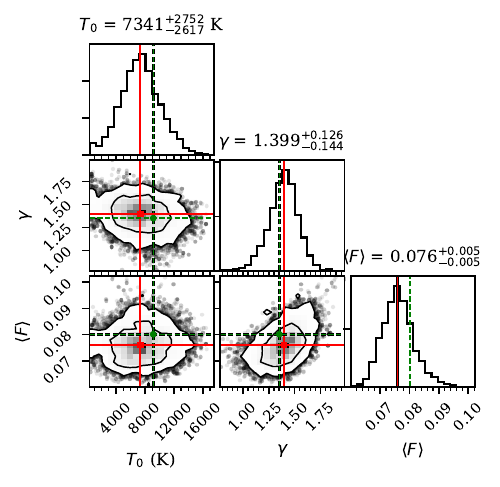} 
    \end{subfigure}
    
    \caption{
        This figure illustrated the results of our inference procedure applied to one mock data set at $z = 5.4$. 
        The top panel shows the resulting models from our inference procedure without re-weighting while the bottom panel has two corner plots that show the resulting parameters, the left without re-weighting and the right with re-weighting. 
        In the top panel, the black points with error bars are the mock data with error bars from the inferred model. 
        The inferred model was calculated by the median (50th percentile) of the MCMC chains of each parameter independently. 
        The inferred model is shown as a red line while the accompanying red text reports errors calculated from the 16th and 84th percentiles of each parameter.
        In comparison, the true model the data was drawn from is the green dotted line and accompanying text. 
        To demonstrate the width of the posterior, multiple faint blue lines are shown which are the models corresponding to the parameters from 100 random draws of the MCMC chain. 
        The bottom left panel shows a corner plot of the values of $T_0$, $\gamma$, and $\langle F \rangle$ that immediately result from our inference procedure. 
        The bottom right panel shows the corner plot of the values of $T_0$, $\gamma$, and $\langle F \rangle$ from our inference procedure using the re-weighting approach. 
        This means the corner plot has been made with the weights calculated from our inference test as described in Section \ref{section: inf re-weight}
    }
    \label{fig:corner_fit}
\end{figure*}

Figure \ref{fig:corner_fit} shows the result of our inference procedure for one mock data set at $z = 5.4$. 
The top panel shows the mock data set with various lines relating to the inference procedure as follows. 
The green dotted line and accompanying text presents the true model that the mock data was drawn from. 
The mock data set is plot as the black points with error bars that come from the diagonal elements of the covariance matrix of the model that is nearest to the inferred model.
The inferred model is the model that comes from the median of each parameter's samples determined independently via the 50th percentile of the MCMC chains. 
The red lines and accompanying text shows the inferred model from MCMC. 
The errors on the inferred model written in the text are the distance between the 16th, 50th, and 84th percentiles of the MCMC chains. 
The blue lines show the models corresponding to 100 random draws from the MCMC chain to visually demonstrate variety of models that come from the resulting posterior. 
The bottom left panel shows a corner plot of the posteriors for $T_0$, $\gamma$, and $\langle F \rangle$.

\subsection{Inference Test and Re-weighting} \label{section: inf re-weight}

We test to check the fidelity of our inference procedure.
This test ensures that the behavior of our posteriors act is statistically correct and checks the validity of any assumptions we make during our inference. 
For example, in this work we used an approximate likelihood in the form of a multivariate Gaussian likelihood.
The \lya forest is known to be a non-Gaussian random field. 
By adopting a multivariate Gaussian likelihood here, we are tacitly assuming that averaging over all pixel pairs when calculating the auto-correlation function will Gaussianize the resulting distribution of the values of the auto-correlation function, as is expected from the central limit theorem. 
We discuss the distribution of these values for our mock data in detail in Appendix \ref{appendix:multi gaussian}. 
If this assumption is not valid our reported errors may be either underestimated or overestimated.

The general idea of our inference test is to compare the true probability contour levels with the ``coverage" probability. 
The coverage probability is the percent of time (over many mock data sets) the true parameters of a mock data set fall within a given probability contour. 
In our case, we compute this over 300 mock data sets where the true parameters considered are sampled from our priors. 
Ideally, this coverage probability should be equal to the chosen probability contour level. 
This calculation can be done at many chosen probabilities resulting in multiple corresponding coverage probabilities.
Existing work that explore this coverage probability include \citet{prangle_2014, ziegel_2014, morrison_2018, sellentin_2019}.

\begin{figure*}
    \centering
    \begin{subfigure}[t]{0.49\textwidth}
        \centering
        \includegraphics[width=\linewidth]{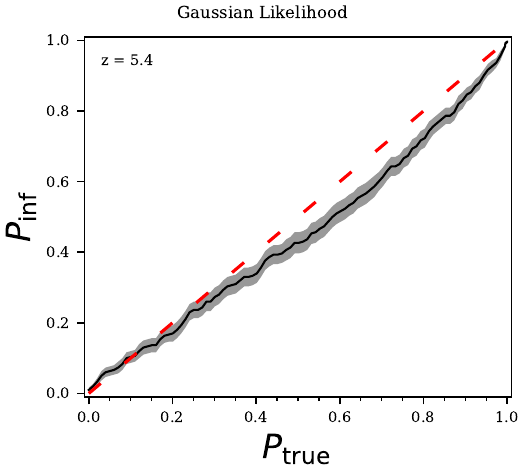} 
    \end{subfigure}
    \hfill
    \begin{subfigure}[t]{0.49\textwidth}
        \centering
        \includegraphics[width=\linewidth]{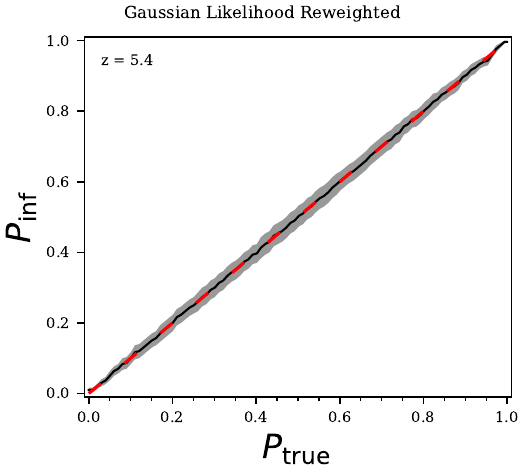} 
    \end{subfigure}
    \caption{
        The left panel of this figure shows the coverage resulting from the inference test from 300 models at $z = 5.4$ drawn from our priors on $T_0$, $\gamma$, and $\langle F \rangle$.
        Here we see that the true parameters for the models fall above the 60th percentile in the MCMC chain $\sim 50\%$ of the time, for example. 
        The right panel of this figure shows the coverage resulting from the inference test with the use of one set of weights to re-weight the posteriors. 
        With these weights we are able to pass the inference test. 
    }
    \label{fig:inf_lines}
\end{figure*}

When considering multiple chosen probabilities, $P_{\text{true}}$, and resulting coverage probabilities, $P_{\text{inf}}$, the results can be plot against each other.
The results of our inference test at $z = 5.4$ from 300 posteriors with true parameters randomly drawn from our priors are shown in the left panel of Figure \ref{fig:inf_lines}.
The grey shaded regions around our resulting line show the Poisson errors for our results.
Again we expect $P_{\text{true}} = P_{\text{inf}}$ which would give the red dashed line in this figure. 
To interpret this plot, first consider one point, for example $P_{\text{true}} \approx 0.6$. 
This represents the 60th percentile contour, which was calculated by the 60th percentile of the probabilities from the draws of the MCMC chain for each mock data set. 
Here, the true parameters fall within the 60th percentile contour only $\approx 52\%$ of the time. 
This implies that our posteriors are too narrow and should be wider such that the true model parameters will fall in the 60th percentile contour more often, so we are in fact underestimating our errors. 
We run this inference test at all $z$ considered in this work and found the deviation from the 1-1 line is worse at higher redshifts. 
See Appendix \ref{appendix:inference z6} for a discussion of the inference test at $z = 6$. 
We additionally run the inference test for mock data generated from a multi-variate Gaussian distribution in Appendix \ref{appendix: gauss data inf}.
The inference test using Gaussian mock data agrees with the 1-1 line, which indicates the reason for the inference test from forward-modeled data failing is that the distribution of the data is not perfectly Gaussian.

There has been much recent work trying to correct posteriors that do not pass this coverage probability test \citep[see e.g.][]{prangle_2014, grunwald_2017, sellentin_2019}.
In this work, we are using the method of Hennawi et al. in prep. where we calculate one set of weights for the MCMC draws that broaden the posteriors in a mathematically rigorous way. 
This method has been described in some detail in \citet{wolfson_2022} so we refer to that paper for details on computing this set of weights. 
For now, we will proceed with discussing the effect of adding these weights to the posteriors.

We show the re-weighted posteriors on $T_0$, $\gamma$, and $\langle F \rangle$ in the bottom right part of Figure \ref{fig:corner_fit}.
The weights give greater importance to values of $T_0$, $\gamma$, and $\langle F \rangle$ that are outside of the $68\%$ contour, effectively broadening the posteriors and increasing the errors on the fit.
For the mock data set in Figure \ref{fig:corner_fit} the re-weighted marginalized posterior for $T_0$ gives $T_0 = 7341^{+2752}_{-2617}$ K whereas before the inferred value was $7282^{+2646}_{-2435}$ K, so the new errors are $\sim 6\%$ larger. 
The reweighted posterior for $\gamma$ gives $\gamma = 1.399^{+0.126}_{-0.144}$ whereas before the inferred value was $1.400^{+0.120}_{-0.132}$, so the new errors are $\sim 7\%$ larger. 
The error on $\langle F \rangle$ does not change. 
When looking at the 2D distributions in this corner plot, such as the $\gamma$, $\langle F \rangle$ distribution on the bottom row middle panel, we can see small regions outside of the main 95\% contour that are highly important. 
This comes from weighting one draw quite highly, which demonstrates how the weights introduce an additional source of noise to the posterior distribution. 

This whole inference procedure is not the optimal and will not give the best possible constraints on $T_0$ or $\gamma$ from the auto-correlation function. 
The need to use re-weighting, or some method to correct our posteriors to pass an inference test, comes from our incorrect (though frequently used) assumption of a multivariate Gaussian likelihood. 
The values of the auto-correlation function at these high $z$ do not sufficiently follow a multivariate Gaussian distribution to justify this assumption, which should be a warning for other studies of the \lya forest at these $z$. 
Using a more correct form of the likelihood (such as a skewed distribution) or likelihood-free inference (such as approximate Bayesian computation as used in \citet{davies_2018_abc} or other machine learning methods) would lead to more optimal posteriors that better reflect the information content of the auto-correlation function.

\subsection{Thermal state measurements} \label{section: thermal results}

We study the distribution of measurements for 100 mock data sets for one ``true" ($T_0$, $\gamma$, $\langle F \rangle$) model in order to account for cosmic variance.
For each $z$ we use the $T_0$, $\gamma$, and $\langle F \rangle$ values reported in Table \ref{tab:central vals}.
Each mock data set is chosen by randomly selecting and averaging the auto-correlation function over 20 skewers. 
For each mock data set, we perform MCMC as described in Section \ref{section: mcmc} and then re-weight the resulting posteriors following Section \ref{section: inf re-weight}.
Once we have the weights and the chains resulting from our inference procedure we can calculate the marginalized posterior for $T_0$ and $\gamma$. 

\begin{figure}
	\includegraphics[width=\columnwidth]{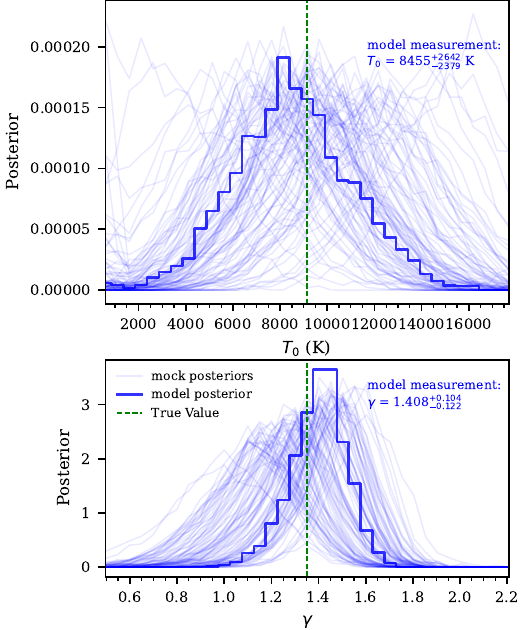}
    \caption{
        This figure shows 100 re-weighted marginalized posteriors of $T_0$ and $\gamma$ at $z = 5.4$ from mock data sets with true $T_0 = 9149$, $\gamma = 1.352$, and $\langle F \rangle = 0.0801$ (faint blue lines).
        The top panel shows the marginalized posteriors for $T_0$ and the bottom panel shows the marginalized posteriors for $\gamma$. 
        It also displays the re-weighted posterior (thick blue histograms) from the model value of the auto-correlation function.
        The measurement resulting from fitting the model like are written in blue text.
        This demonstrates the different possible behaviors the posterior can have for different mock data sets with the same ``true" $T_0$, $\gamma$, and $\langle F \rangle$ values. 
    }
    \label{fig:many_posterior_temp_gamma}
\end{figure}

At $z = 5.4$, all 100 marginalized re-weighted posteriors are shown as the faint blue lines in Figure \ref{fig:many_posterior_temp_gamma} for $T_0$ (top panel) and $\gamma$ (bottom panel).
Attempting to fit the model value of the auto-correlation function removes the luck of the draw that exists in selecting mock data and gives the optimal precision of the posteriors. 
The resulting posteriors from fitting the model is shown as the thick blue histogram in the figure. 
The measurement resulting from this fit is written in the blue text of this figure and the values at each $z$ are reported in Table \ref{tab:measurements}. 

The re-weighted histograms in Figure \ref{fig:many_posterior_temp_gamma} are noisy, much like is seen in the bottom right panel of Figure \ref{fig:corner_fit}. 
This is a direct consequence of our re-weighting procedure and will be improved with further work on likelihood-free inference. 
For $T_0$ the model value of the auto-correlation function gives a posterior with a width that is typical of those from the mock data.
For $\gamma$ the posterior has a slightly narrower peak. 
Also for $\gamma$, posteriors that peak at lower $\gamma$ values are broader than those that peak at higher $\gamma$ values. 
Both model posteriors contain the true value of $T_0$ and $\gamma$ within their $1 \sigma$ error bars.

Table $\ref{tab:measurements}$ reports the measurements that result from using the model values of the auto-correlation function as our data at all $z$. 
This is an ideal scenario that removes luck of the draw from the resulting measurement. 
The first (third) column contains the ``true" modeled value of $T_0$ ($\gamma$) at each $z$ that was used in this measurement. 
The second (fourth) column contains the measurements for $T_0$ ($\gamma$) calculated by the 16th, 50th, and 84th percentiles. 
In general the trend of the errors is to increase with increasing redshift. 
At $z = 5.4$, the measurement of the model constrains $T_0$ to 29\% and $\gamma$ to 9\%.

\begin{table}
    \begin{tabular}{|ccccc|}
        \hline
        $z$ & Model $T_0$ & Measured $T_0$         & Model $\gamma$ & Measured $\gamma$         \\ \hline
        5.4 & 9149        & $8455^{+2642}_{-2379}$ & 1.352          & $1.408^{+0.104}_{-0.122}$ \\
        5.5 & 9354        & $8643^{+3152}_{-3054}$ & 1.338          & $1.422^{+0.116}_{-0.141}$ \\
        5.6 & 9572        & $8480^{+3720}_{-3642}$ & 1.324          & $1.433^{+0.121}_{-0.151}$ \\
        5.7 & 9804        & $8222^{+5188}_{-4176}$ & 1.309          & $1.460^{+0.139}_{-0.166}$ \\
        5.8 & 10050       & $8346^{+4926}_{-4576}$ & 1.294          & $1.485^{+0.157}_{-0.204}$ \\
        5.9 & 10320       & $7892^{+6111}_{-4655}$ & 1.278          & $1.513^{+0.170}_{-0.223}$ \\
        6.0 & 10600       & $9574^{+6219}_{-5133}$ & 1.262          & $1.511^{+0.196}_{-0.256}$ \\ \hline
    \end{tabular}
    \centering
    \caption{
        This table contains the results of analyzing the $T_0$ and $\gamma$ posteriors for the model value of the auto-correlation function.
        The first (third) column contains the modeled value of $T_0$ ($\gamma$) at each $z$ that was used in this measurement. 
        The second (fourth) column contains the measurements for $T_0$ ($\gamma$) calculated by the 16th, 50th, and 84th percentiles. 
        In general the trend of the errors is to increase with increasing redshift. 
        }
    \label{tab:measurements}
\end{table}

\begin{figure}
	\includegraphics[width=\columnwidth]{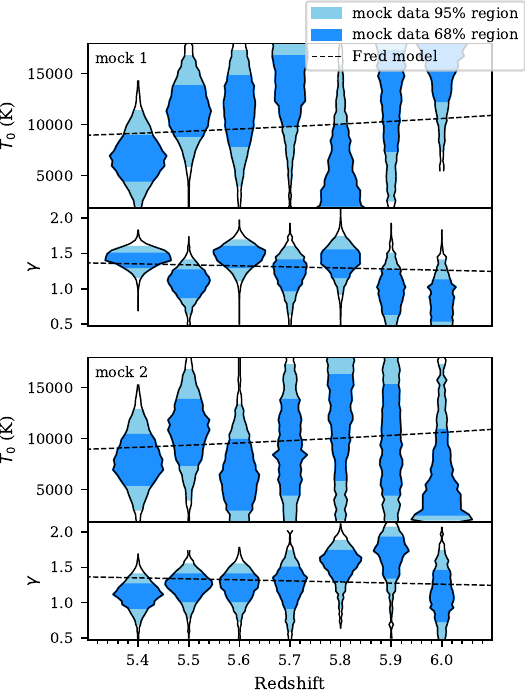}
    \caption{
        The two sections of this figure show the marginalized posteriors for a random mock data set at each $z$ for both $T_0$ and $\gamma$. 
        The only difference between the upper and lower sections of the figure are the random mock data set selected. 
        The top panel in both sections shows the marginalized posteriors for $T_0$ while the bottom sections show the same for $\gamma$ from the same data set. 
        For each posterior, the light blue shaded region demarcates the 2.5th and 97.5th percentile of the weighted MCMC draws while the darker blue shaded region demarcates the 17th and 83rd percentile of the weighted MCMC draws. 
        The behavior of each posterior at the different $z$ is determined by the luck of the draw when selecting the mock data. 
        The black dot dashed line shows the true model values from in Figure \ref{fig:temp_gamma_model_evolution} which are reported in Table \ref{tab:central vals}.    
        }
    \label{fig:temp_gamma_violin_plot}
\end{figure}

In order to visualize the differences between measurements at different redshifts, we plot the results for two random mock data sets in the two sections of Figure \ref{fig:temp_gamma_violin_plot}. 
For each section, the top panel shows the marginalized posteriors for $T_0$ while the bottom shows the marginalized posteriors for $\gamma$.
Each violin is the re-weighted marginalized posterior for one randomly selected mock data set at the corresponding redshift. 
The light blue shaded region demarcates the 2.5th and 97.5th percentiles (2$\sigma$) of the MCMC draws while the darker blue shaded region demarcates the 16th and 84th percentiles (1$\sigma$) of the MCMC draws. 
The dot dashed line is the true simulated model value evolution as shown in Figure \ref{fig:temp_gamma_model_evolution} and reported in Table \ref{tab:central vals}. 

Looking at the posteriors for a given redshift (one column in the figure), the only difference between the posteriors is the random mock data set drawn. 
This still produces different precision results as seen in Figure \ref{fig:many_posterior_temp_gamma} for $z = 5.4$. 
For the different posteriors within one section of this figure, the random mock data set differs as does the true values of $T_0$ and $\gamma$.
Again, the individual posteriors are noisy, resulting from the re-weighting procedure as described in Section \ref{section: inf re-weight}.  
The behavior here echos that found with the model measurements where the precision of the constraints on $T_0$ and $\gamma$ decrease with increasing $z$. 
In the highest redshift bins, $z > 5.7$, the posteriors for the mock data sets have high values at the boundary of our prior much more often.

\section{Inhomogeneous Reionization} \label{section: inhomo reion}

So far in this work we have used a semi-numerical method to ``paint on" different thermal states to our simulations for a tight temperature-density relationship. 
This is sufficient to explore the sensitivity of the \lya forest flux auto-correlation function to the thermal state of the IGM at high-redshifts. 
However, as previously discussed, recent measurements of the \lya optical depth at $z > 5.5$ have shown scatter that can't be explained by density fluctuations alone \citep{fan_2006, becker_2015, bosman_2018, eilers_2018, bosman_2021_data}. 
It is possible that these fluctuations come from fluctuations in the temperature field \citep{daloisio_2015, davies_2018_thermal} or fluctuations in the UVB \citep{davies_furlanetto_2016, gnedin_2017, daloisio_2018}. 
Fluctuations in either of these fields can arise if reionization is extended or patchy.

On top of the measurements of fluctuations in the \lya forest optical depth at $z > 5.5$, recent measurements of the mean free path of ionizing photons at $z > 5$ suggest a UVB that cannot be well described by uniform fields \citep{becker_2021, bosman_2021_limit, gaikwad_2023, zhu_2023}. 

In order to explore the effect of temperature and UVB fluctuations on the \lya forest flux auto-correlation function, we consider a set of four simulation models. 
These simulations have two different reionization models (one of which causes temperature fluctuations) and two UVB models (one of which has fluctuations).
These simulations and their results will be described in detail in the following sections.

\subsection{Simulation box}

\begin{figure}
	\includegraphics[width=\columnwidth]{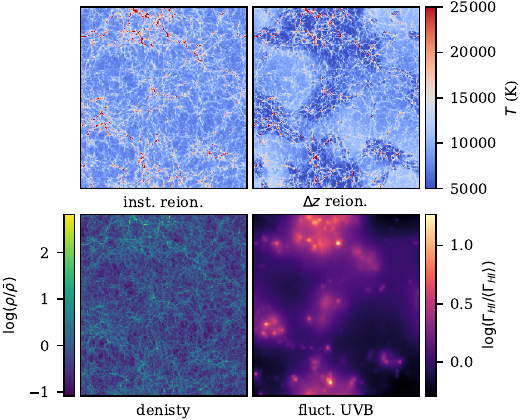}
    \caption{
        This figure shows slices through the simulation boxes of the density (bottom left), the temperature field (top row), and UVB (bottom right) for the \texttt{Nyx} simulation described in Section \ref{section: inhomo reion}.
        On the left in the top panel is the temperature field for the instantaneous reionization model while the right shows the inhomogeneous reionization model. 
        With inhomogeneous reionization, there is a greater scatter in the temperature field, as can be seen by the greater abundance of colder (darker blue) regions. 
        These cold regions correspond to the regions of higher density in the bottom left panel.
        The bottom right panel shows a slice through the UVB field of the simulation with $\lambda_{\text{mfp}}$ = 15 cMpc, which gives a fluctuating UVB. 
        The largest UVB values are in the same location as the high density areas shown in the bottom left panel. 
        These are the densest regions of the simulation which contain the majority of the sources of ionizing photons. 
    }
    \label{fig:thermal_uvb_slices}
\end{figure}

For these models we use an additional \texttt{Nyx} simulation box with a size of $L_{\text{box}} = 40$ cMpc h$^{-1}$ and $2048^3$ resolution elements at $z = 5.8$. 
A slice through the density field of this simulation is shown in the bottom left panel of Figure \ref{fig:thermal_uvb_slices}.

The two different reionization models considered in this simulation are either instantaneous reionization (inst. reion.) or an extended, inhomogeneous model ($\Delta z$ reion.).
The first model of reionization (inst. reion.) is the ``flash" method described in \citet{onorbe_2019}. 
This method assigns all resolution elements the same redshift of reionization where the heat is injected. 
It is possible for cells to be ionized before the redshift through other processes such as collisional reionization. 
We use $z^{\text{median}}_{\text{reion, HI}} = 7.75$.
The second model of reionization ($\Delta z$ reion.) is the inhomogeneous reionization described in \citet{onorbe_2019}. 
To summarize this method, reionization is implemented by assigning each resolution element a redshift of reionization and then injecting some amount of energy, $\Delta T$, at that resolution element and redshift. 
We use $z^{\text{median}}_{\text{reion, HI}} = 7.75$ and $\Delta z_{\text{reion, HI}} = 4.82$.
In both of these reionization model the energy injected is $\Delta T = 2 \times 10^{4}$ K. 
The top row of Figure \ref{fig:thermal_uvb_slices} shows slices through the resulting temperature field from these two simulations: inst. reion. on the left and $\Delta z$ reion. on the right. 
From this figure we see that the inhomogeneous, $\Delta z$ reion. model has a larger scatter in the temperature with the greater abundance of colder (darker blue) regions. 
These cold regions correspond to the regions of higher density in the bottom left panel.
This follows from the model of reionization where the denser regions reionize (and are heated) first and thus have more time to cool to a lower temperature by $z = 5.8$.

In addition to a constant, uniform UVB model (const. UVB), we have an inhomogeneous, fluctuating UVB model (fluct. UVB) using the same method as that presented in \citet{onorbe_2019} with $\lambda_{\text{mfp}} = 15$ cMpc.
This follows the approach of \citet{davies_furlanetto_2016} where we consider modulations in the ionization state of optically thick absorbers assuming that $\lambda_{\text{mfp}} \propto \Gamma_{\text{UVB}}^{2/3} / \Delta$ where $\Delta$ is the local matter density.
For the fluctuating UVB, \guvb was calculated on a uniform grid of $64^3$ at $z = 6$ and then linearly interpolated the $\log\Gamma_{\text{UVB}}$ field to match the hydrodynamical simulation with $2048^3$. 
The bottom right panel of Figure \ref{fig:thermal_uvb_slices} shows a slice through the fluct. UVB model. 
The largest UVB values are in the same location as the high density areas shown in the bottom left panel. 
These are the densest regions of the simulation which contain the majority of the sources of ionizing photons.

The four resulting models consist of (1) inst. reion., const. UVB (2) inst. reion., fluct. UVB, (3) $\Delta z$ reion., const. UVB and (4) $\Delta z$ reion., fluct. UVB. 
All four models are normalized to $\langle F \rangle = .0172$, which was the measured value at $z = 5.8$ from \citet{bosman_2021_data}.
We do not consider multiple values of $\langle F \rangle$ for these models since they represent four discrete models and we will not try to constrain any parameters.

\begin{figure*}
	\includegraphics[width=2\columnwidth]{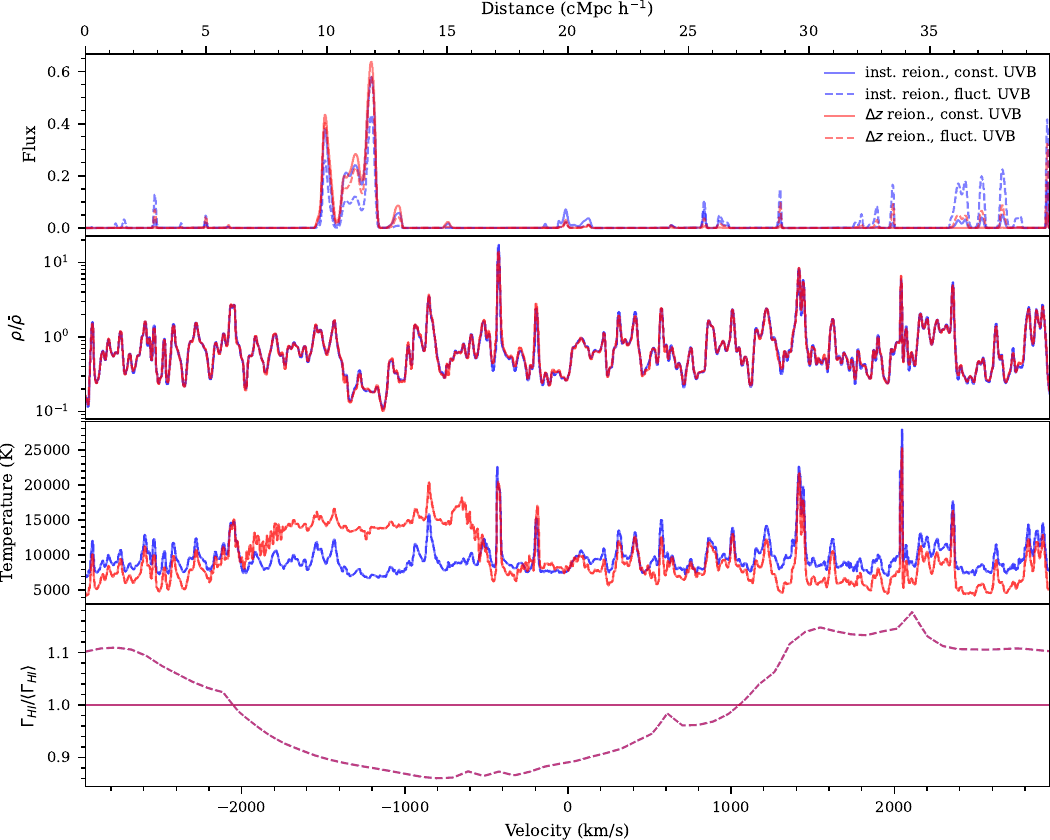}
    \caption{
        This figure shows one skewer from the four various reionization models at $z = 5.8$.
        The top panel shows the resulting \lya forest flux. 
        The second panel shows the density field along the skewer. 
        The third panel shows the temperature along the skewer. 
        The bottom panel shows the UVB background values. 
        Each panel has four lines representing the inst. reion., const. UVB (solid blue), inst. reion., fluct. UVB (dashed blue), $\Delta z$ reion., const. UVB (solid red), and $\Delta z$ reion., fluct. UVB (dashed red) models. 
        Comparing the solid lines to each other isolates the effect of temperature fluctuations only. 
        When comparing these two models, we see that a positive scatter in the temperature of the IGM leads to increased flux over $\SI{-1600}{\kilo\meter\per\second} < v < \SI{-1000}{\kilo\meter\per\second}$. 
        Comparing the dashed lines to the solid lines of the same color isolates the effect of UVB fluctuations on the flux of the \lya forest. 
        For example consider $v > \SI{1000}{\kilo\meter\per\second}$ where the fluct. UVB (dashed) models in the bottom panel are constantly greater than the const. UVB (solid) models.
        For the corresponding region in the top panel, the flux is boosted for the fluct. UVB (dashed) models. 
    }
    \label{fig:flux_four_model}
\end{figure*}

We now consider the effect of these four simulation models on the transmitted flux.
Figure \ref{fig:flux_four_model} shows one skewer from each of the four different reionization models at $z = 5.8$. 
The top panel shows the resulting \lya forest flux. 
The second panel shows the density field along the skewer. 
The third panel shows the temperature along the skewer. 
The bottom panel shows the UVB background values. 
Each panel has four lines representing the inst. reion., const. UVB (solid blue), inst. reion., fluct. UVB (dashed blue), $\Delta z$ reion., const. UVB (solid red), and $\Delta z$ reion., fluct. UVB (dashed red) models. 
Comparing the solid lines to each other isolates the effect of temperature fluctuations only. 
When comparing these two models, we see that a positive scatter in the temperature of the IGM leads to increased flux over $\SI{-1600}{\kilo\meter\per\second} < v < \SI{-1000}{\kilo\meter\per\second}$. 
Comparing the dashed lines to the solid lines of the same color isolates the effect of UVB fluctuations on the flux of the \lya forest. 
For example consider $v > \SI{1000}{\kilo\meter\per\second}$ where the fluct. UVB (dashed) models in the bottom panel are constantly greater than the const. UVB (solid) models.
For the corresponding region in the top panel, the flux is boosted for the fluct. UVB (dashed) models.

In general for these models, the UVB variations are anti-correlated with the temperature fluctuations.
This follows from the dense regions in the simulations causing negative temperature fluctuation and positive UVB simulation as discussed earlier. 
For example, consider the positive temperature fluctuation and negative UVB fluctuation at $\SI{-1600}{\kilo\meter\per\second} < v < \SI{-1000}{\kilo\meter\per\second}$. 
Overall this anti-correlation will result in the effects of these two fluctuating fields to cancel out, causing the $\Delta z$ reion., fluct. UVB (dashed red) model to look similar to the inst. reion., const. UVB (solid blue) model.
This is indeed generally seen across the flux panel of Figure \ref{fig:flux_four_model}.

From here, we forward model the skewers in the same way as discussed in Section \ref{section: forward modeling} with $R = 30000$ and $\snr = 30$. 
The only difference is that we leave the skewers with the full 40 cMpc h$^{-1}$ length and then use only 15 (where before we used 20) skewers when calculating mock data sets. 
The mock data sets here and in the previous section contain the same path length corresponding to 20 observed quasars with $\Delta z$ = 0.1. 
We do not show an example of the forward modeled skewer here as they are very similar to that shown in Figure \ref{fig:flux_noise}.

\subsection{Auto-correlation}

The auto-correlation function is computed via Equation \eqref{eq:autocorr} and the covariance matrices are computed via Equation \eqref{eq:covariance}. 

\begin{figure*}
	\includegraphics[width=2\columnwidth]{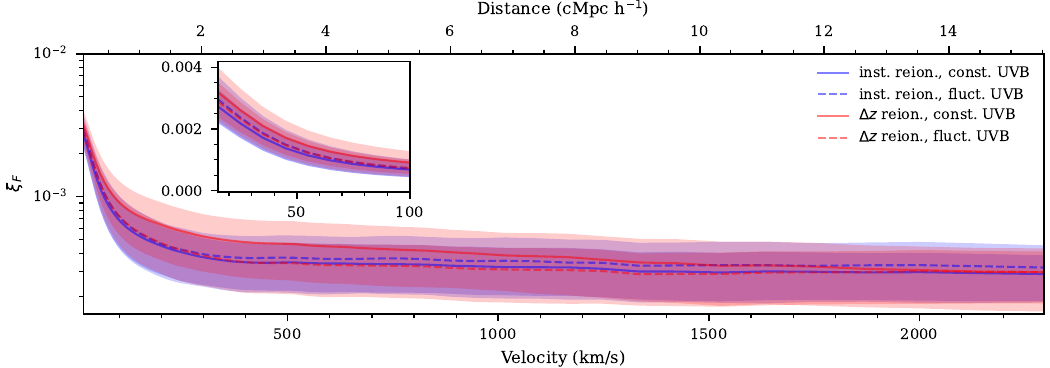}
    \caption{
        This figure shows the correlation function for the four reionization models at $z = 5.8$ with a logarithmic y-axis. 
        The lines show the model values of the correlation function while the shaded region shows the errors estimated from the diagonal of the covariance matrices. 
        The colors and lines tyles here match those in Figure \ref{fig:flux_four_model} with the inst. reion., const. UVB (solid blue), inst. reion., fluct. UVB (dashed blue), $\Delta z$ reion., const. UVB (solid red), and $\Delta z$ reion., fluct. UVB (dashed red) models. 
        Comparing the solid lines isolates the effect of temperature fluctuations while comparing the same colors isolates the effect of UVB fluctuations. 
        Note that the shaded regions are about the same size for all four models. 
        The inset shows the first $\SI{100}{\kilo\meter\per\second}$ of the auto-correlation functions with a linear y-axis to see the differences at small scales. 
    }
    \label{fig:correlation_four_model}
\end{figure*}

Figure \ref{fig:correlation_four_model} shows the correlation function for the four reionization models at $z = 5.8$ with a logarithmic y-axis. 
The inset shows the first $\SI{100}{\kilo\meter\per\second}$ of the auto-correlation functions with a linear y-axis to highlight the differences at small scales. 
The lines show the model value while the shaded regions are the error estimated from the diagonals of the covariance matrices. 
The colors and line styles here match those in Figure \ref{fig:flux_four_model} with the inst. reion., const. UVB (solid blue), inst. reion., fluct. UVB (dashed blue), $\Delta z$ reion., const. UVB (solid red), and $\Delta z$ reion., fluct. UVB (dashed red) models. 
Comparing the same line styles isolates the effect of temperature fluctuations while comparing the same colors isolates the effect of UVB fluctuations. 
Note that the shaded regions are about the same size for all four models. 

First compare inst. reion., const. UVB (solid blue) and $\Delta z$ reion., const. UVB (solid red), which isolates the effect of temperature fluctuations. 
The model values for these models show that adding temperature fluctuations boosts the value of the auto-correlation function for $\Delta v < \SI{1800}{\kilo\meter\per\second}$. 
This follows from the additional variation added by the temperature fluctuations. 

Now consider inst. reion., const. UVB (solid blue) and inst. reion., fluct. UVB (dashed blue), which adds UVB fluctuations to a model without temperature fluctuations. 
Comparing these line in the inset shows that adding UVB fluctuations increases the value of the auto-correlation function on small scales. 
This result falls in line with that found in \citet{wolfson_2022} which says that a shorter $\lambda_{\text{mfp}}$ value leads to greater boosts on small scales of the auto-correlation function. 
At larger scales there is a slight boost in the inst. reion., fluct. UVB (dashed blue) seen with the logarithmic scale.

Finally consider $\Delta z$ reion., const. UVB (solid red) and $\Delta z$ reion., fluct. UVB (dashed red), which compares adding UVB fluctuations to a model with temperature fluctuations. 
In this case adding UVB fluctuations decreases the value of the auto-correlation function for $\Delta v < \SI{1800}{\kilo\meter\per\second}$. 
This is the opposite effect as adding UVB fluctuations to a model without temperature fluctuations (seen in comparing the blue lines) and the results from \citet{wolfson_2022}. 
However, this results from the anti-correlation between the UVB and temperature fluctuations resulting from the correlations with the density field.
For a fluctuating UVB, the UVB is highest where the density is greatest as this is where sources are.
However, for a fluctuating temperature model, the temperature is lowest where the density is greatest, which would decrease the transmitted flux. 
This causes more constant flux levels and decreases the auto-correlation function values at these small scales, as seen in these lines. 
Ultimately, the correlations with density cause the $\Delta z$ reion., fluct. UVB (dashed red) to be most similar to the inst. reion., const. UVB model (solid blue). 
Note that on small scales there is still a boost in the $\Delta z$ reion., fluct. UVB model (dashed red) over the inst. reion., const. UVB model model (solid blue), which comes from increased variation in the flux.

\subsection{Ruling-out Reionization scenarios} \label{sec:ruling out}

For these four reionization models, there is no grid of parameters that can be constrained via MCMC. 
Instead, we will investigate how confidently other models can be ruled out given mock data from a single model. 
We will rule out models via the likelihood ratio, $\mathcal{R}$, which is defined as 
\begin{equation}
    \mathcal{R} = \frac{\mathcal{L}(\text{model})}{\mathcal{L}(\text{reference model})}
    \label{eq: like ratio}
\end{equation}
Again for this we assume the likelihood, $\mathcal{L}$, is the multivariate Gaussian likelihood from Equation \eqref{eq:gauss_like}.

Here we assume that the mock data comes from the $\Delta z$ reion., fluct. UVB model (red dashed lines in the Figures \ref{fig:flux_four_model} and \ref{fig:correlation_four_model}). 
Therefore, we will be looking at the value of the likelihood for the mock data sets using the other three reionization models divided by the likelihood for what we know is the true mock data model ($\Delta z$ reion., fluct.). 
To investigate the distribution of potential likelihood ratio values, we use 1000 mock data sets.

\begin{figure}
	\includegraphics[width=\columnwidth]{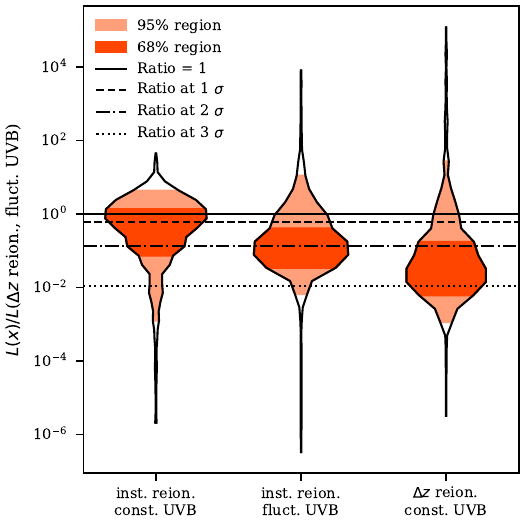}
    \caption{
        This figure shows the distribution of likelihood ratios from 1000 mock data sets where the model the mock data originates from is the $\Delta z$ reion., fluct. UVB model.  
        The violin plots show the full distribution where the light orange shaded region demarcates the 2.5th and 97.5th percentiles (2$\sigma$) of the ratio values while the darker orange shaded region demarcates the 16th and 84th percentiles (1$\sigma$) of the ratio values. 
        The solid black line shows where the ratio is equal to 1, which is where both models are just as likely given the mock data. 
        The dashed, dot-dashed, and dotted back lines show the value where the alternative models are ruled out at the 1, 2, and 3 $\sigma$ levels respectively. 
    }
    \label{fig:likelihood_ratio_four_models}
\end{figure}

The distribution of the 1000 likelihood ratio values for each of the alternative reionization models are shown in Figure \ref{fig:likelihood_ratio_four_models}. 
The violin plots show the full distribution where the light orange shaded region demarcates the 2.5th and 97.5th percentiles (2$\sigma$) of the ratio values while the darker orange shaded region demarcates the 16th and 84th percentiles (1$\sigma$) of the ratio values. 
The solid black line shows where the ratio is equal to 1, which is where both models are just as likely given the mock data. 
The dashed, dot-dashed, and dotted back lines show the value where the alternative models are ruled out at the 1, 2, and 3 $\sigma$ levels respectively. 

Overall, it is most difficult to rule out the inst. reion., const. UVB model (solid blue lines in previous plots) from the $\Delta z$ reion., fluct. UVB model, as is seen in the left most violin in Figure \ref{fig:likelihood_ratio_four_models}. 
This distribution has 46.9\% of the mock data sets that favor the incorrect, alternative reionization scenario than the true $\Delta z$ reion., fluct. UVB model. 
Then only 50.4\%, 24.3\%, and 7.6\% of mock data sets can be ruled out at the 1, 2, and 3 $\sigma$ levels respectively. 
This follows from the auto-correlation values for these models seen in Figure \ref{fig:correlation_four_model} and the discussion there about how the temperature fluctuations and UVB fluctuations are anti-correlated and thus produce an auto-correlation function most similar to the inst. reion., const. UVB model which lacks both of these effects. 

The next most difficult model to rule out is arguably the inst. reion., fluct. UVB model (dashed blue lines in the previous plot) as seen in the central violin in Figure \ref{fig:likelihood_ratio_four_models}. 
This distribution has only 17.6\% of the mock data sets that favor the incorrect, alternative reionization scenario than the true $\Delta z$ reion., fluct. UVB model. 
Then 74.8\%, 46.3\%, and only 5.2\% of mock data sets can be ruled out at the 1, 2, and 3 $\sigma$ levels respectively. 
Between this and the left plot there are fewer mock data sets here that can be ruled out at least at the 3$\sigma$ level but almost half of them can be ruled out at 2$\sigma$. 

The easiest model to rule out is finally the $\Delta z$ reion., const. UVB model (solid red lines in the previous plots) as seen in the right most violin in Figure \ref{fig:likelihood_ratio_four_models}. 
This distribution has only 12.5\% of the mock data sets that favor the incorrect, alternative reionization scenario than the true $\Delta z$ reion., fluct. UVB model. 
Then 84.2\%, 68.6\%, and 22.1\% of mock data sets can be ruled out at the 1, 2, and 3 $\sigma$ levels respectively, which is the greatest percentages out of the three alternative models. 
This also follows from the differences between these models in Figure \ref{fig:correlation_four_model}. 
The $\Delta z$ reion., const. UVB model has the greatest values of the auto-correlation function at most scales, making it the easiest to distinguish. 

Again, we are looking at the distribution of the likelihood for 1000 mock data sets here and luck of the draw would ultimately determine if it is possible to rule out each model with a given observational data set. 
For each alternative model it is possible that the incorrect models is favored over the true model from which the mock data was drawn, though this was always true for less than half of the mock data sets.

\section{Conclusions} \label{section: conclusion}

In this work we have investigated the precision of possible constraints on the thermal state of the IGM from the auto-correlation function of \lya forest flux in high resolution quasar observations. 
This came in two forms: constraining $T_0$ and $\gamma$ when the IGM thermal state follows a tight power law of the form of Equation \eqref{eq:temperature-density relation} and models with temperature fluctuations from different reionization scenarios. 

We discussed the results of constraints on $T_0$ and $\gamma$ in Section \ref{section: methods}. 
Overall, we found that the auto-correlation function is sensitive to $T_0$ and $\gamma$ across multiple redshift bins for realistic mock data of 20 quasars with $R = 30000$. 
We computed the marginalized re-weighted posterior for $T_0$ and $\gamma$ for 100 mock data sets at $5.4 \leq z \leq 6.0$. 
The re-weighted posterior showed a variety of behaviors based on the luck of the draw of the mock data chosen, the true value of $\gamma$ for the mock data, and the data set size at each $z$. 
We considered an ideal data set which had the model value of the auto-correlation function, effectively removing the luck of the draw from our measurement.
The error on these measurements for both the $T_0$ and $\gamma$ increase with redshift, which may be from the low $\langle F \rangle$. 
At $z = 5.4$ we found that ideal data can constrain $T_0$ to 29\% and $\gamma$ to 9\%.

Note that our procedure uses a multi-variate Gaussian likelihood, MCMC, and a set of weights for the MCMC chains that ensures our posteriors pass an inference test.
The original failure of our procedure to pass an inference test is due to the incorrect assumption that the auto-correlation function follows a multi-variate Gaussian distribution, as discussed in Appendix \ref{appendix:multi gaussian}. 
This result should caution against using a multi-variate Gaussian likelihood with other statistics, such as the power spectrum, when making measurements at $z > 5$ as the same issue of non-Gaussian data likely applies. 
In the future, better likelihoods or likelihood-free inference will allow for a more optimal inference procedure \citep[see e.g.][]{davies_2018_abc,alsing_2019}.

We discussed the results of measurements for four different reionization models in Section \ref{section: inhomo reion}.  
For these models we don't have a grid where we could constrain the physical parameters. 
Instead, we quantitatively investigate these models by computing the likelihood ratio between two models, assuming a Gaussian distribution of data. 
Looking at a model which has temperature fluctuations and UVB fluctuations, we found that it is easiest to distinguish between this and the model with temperature fluctuations and no UVB fluctuations and it is most difficult to distinguish between this and the model with no temperature or UVB fluctuations. 
The actual ability to distinguish between models depends on the luck of the draw for the actual data that is measured. 
We found that 50.4\% of mock data sets from the model with temperaure and UVB fluctuations can rule out a model without temperature or UVB fluctuations at $>1\sigma$ level.

Note that the UVB models used in this section were computed in in a small box (40 cMpc h$^{-1}$) which suppresses UVB fluctuations on all scales, as was discussed in \citet{wolfson_2022}. 
Suppressing fluctuations in the UVB causes the auto-correlation signal to be lower in these boxes.
For this reason, it may be easier to distinguish between models with and without UVB fluctuations if they were generated in a larger box. 
Thus, future work on UVB models will be necessary to get the best possible constraints on reionization from these models. 

Both the thermal state and the UVB fluctuations affect the \lya forest flux auto-correlation function. 
Modeling both of these physical effects by varying multiple parameters in a larger box will allow the auto-correlation function to constrain the two simultaneously. 
This will allow us to put quantitative constraints on the thermal state of the IGM, the $\lambda_{\text{mfp}}$ that describes the UVB, and ultimately reionization. 
We leave this exploration to future work.

Constraining the thermal state of the IGM, such as through constraining characteristic $T_0$ and $\gamma$ values at high-$z$ is an important method to constrain reionization. 
Measuring these parameters at $z > 5$ is a difficult task that has so far been done with few methods \citep{boera_2019, walther_2019, gaikwad_2020}. 
This work has shown that the auto-correlation function of the \lya forest flux provides a new, competitive way to constrain $T_0$ and $\gamma$ in multiple redshift bins at $z \geq 5.4$.

\section*{Acknowledgements}

We acknowledge helpful conversations with the ENIGMA group at UC Santa Barbara and Leiden University. 
JFH acknowledges support from the European Research Council (ERC) under the European Union’s Horizon 2020 research and innovation program (grant agreement No 885301) and from the National Science Foundation under Grant No. 1816006.

This research used resources of the National Energy Research Scientific Computing Center (NERSC), a U.S. Department of Energy Office of Science User Facility located at Lawrence Berkeley National Laboratory, operated under Contract No. DE-AC02-05CH11231.

%%%%%%%%%%%%%%%%%%%%%%%%%%%%%%%%%%%%%%%%%%%%%%%%%%
\section*{Data Availability}

The simulation data analyzed in this article will be shared on reasonable request to the corresponding author.



\bibliographystyle{mnras}
\bibliography{tex_doc_corr_thermal} % if your bibtex file is called example.bib

%\begin{thebibliography}{99}
%\end{thebibliography}

%%%%%%%%%%%%%%%%%%%%%%%%%%%%%%%%%%%%%%%%%%%%%%%%%%

%%%%%%%%%%%%%%%%% APPENDICES %%%%%%%%%%%%%%%%%%%%%

\appendix

\section{Power spectra models} \label{appendix:power_cos}

\begin{figure*}
	\includegraphics[width=2\columnwidth]{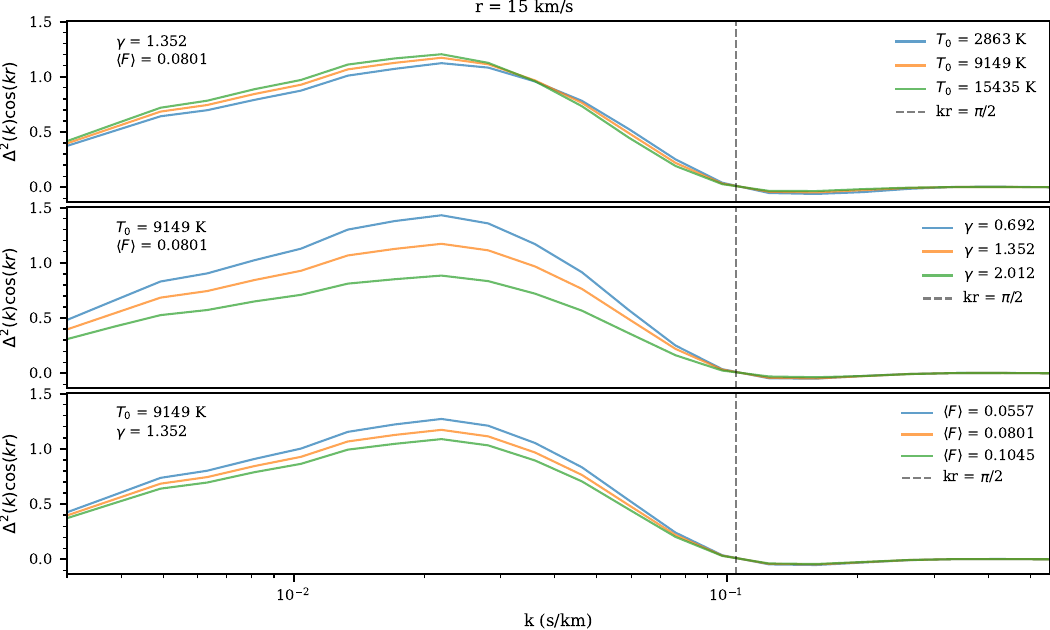}
    \caption{
        This figure shows the mean value of $\Delta^2\cos(kr)$ where $r = \SI{15}{\kilo\meter\per\second}$ for different sets of parameters. 
        Each panels varies one parameter while keeping the others constant with $T_0$, $\gamma$, and $\langle F \rangle$ varying in the top, middle, and bottom panels respectively.
        For these panels the solid lines show the model values calculated by averaging $\Delta^2$ from all forward modeled skewers available. 
        This figure is meant to explain the behavior of the auto-correlation seen in Figure \ref{fig:correlation_three_panels} at $\Delta v = r = \SI{15}{\kilo\meter\per\second}$ due to the relation in Equation \eqref{eq:autocorr_int_power}.
    }
    \label{fig:app_power_cos}
\end{figure*}

As explained in Section \ref{sec:autocorr}, the dimensionless power, $\Delta^2_{\delta_f}(k)$, can be written as the Fourier transform of the auto-correlation function of the flux contrast, $\xi_{\delta_f}(\Delta v)$. 
$\xi_{\delta_f}(\Delta v)$ is explicitly written in terms of $\Delta^2_{\delta_f}(k)$ in Equation \eqref{eq:autocorr_int_power}, which says $\xi_{\delta_f}(\Delta v)$ is the integral of $\Delta^2_{\delta_f}(k)\cos(k\Delta v)$ in logarithmic $k$ bins. 
We refer to $\Delta^2_{\delta_f}(k)\cos(k\Delta v)$ as the integrand for the rest of this discussion.
To build intuition for the auto-correlation function at small scales we show the integrand for $\Delta v = r = \SI{15}{\kilo\meter\per\second}$ in Figure \ref{fig:app_power_cos}. 

This Figure mimics the set up of Figure \ref{fig:correlation_three_panels} for the auto-correlation function where each panels varies one parameter while keeping the others constant.
For these panels the solid lines show the model values calculated by averaging $\Delta^2$ from all forward modeled skewers available. 
The vertical grey dashed line shows where $\cos(kr) = 0$.

In the top panel $T_0$ varies while $\gamma$ and $\langle F \rangle$ are constant. 
At small $k$ the larger values of $T_0$ have larger values of the integrand while at small $k$ there is thermal cutoff and smaller values of $T_0$ now have larger values of the integrand. 
When integrating over these logarithmic bins the greater $T_0$ values end up with more area and thus the auto-correlation functions are also greater.

\section{Convergence of the Covariance Matrices} \label{appendix:converge}

\begin{figure}
	\includegraphics[width=\columnwidth]{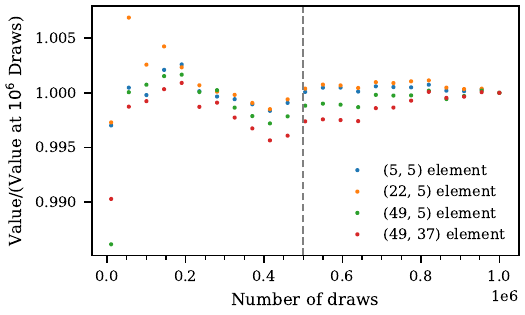}
    \caption{
        This figure shows the behavior of four elements of the model covariance matrix ($z = 5.4$, $T_0 = 9149$ K, $\gamma = 1.352$, and $\langle F \rangle = 0.0801$) for different numbers of mock draws. 
        At all values of the number of mocks considered, the covariance elements fall within 2\% of their final value. 
        By around $\sim 100000$ draws, all of the values fall within 0.5\% of the final value. 
        For this reason, using 500000 mock draws is sufficient to generate the covariance matrices used in this study. 
        500000 mock draws is represented by the vertical dashed grey line. 
    }
    \label{fig:covar_conv}
\end{figure}

We calculate the covariance matrices for our models with mock draws, as defined in equation \eqref{eq:covariance}.
Using mock draws is inherently noisy and it should converge as $1 / \sqrt{N}$ where $N$ is the number of draws used. 
As stated in the text, we used 500000 mock draws. 
To check that this number is sufficient to minimize the error in our calculation, we looked at the behavior of elements of one covariance matrix in Figure \ref{fig:covar_conv}. 
This covariance matrix is for the model with $z = 5.4$, $T_0 = 9149$ K, $\gamma = 1.352$, and $\langle F \rangle = 0.0801$, which is the ``true'' model at this redshift. 
The correlation matrix for this model is also shown in Figure \ref{fig:covar_correlation}. 

The values in the plot have been normalized to 1 at $10^6$ draws. 
The four elements have been chosen such that there is one diagonal value and three off-diagonal values in different regions of the matrix. 
At all values of the number of mock draws considered, the covariance elements fall within 2\% of their final value. 
By around $\sim 100000$ draws, the values fall within 0.5\% of the final value. 
For this reason, using 500000 mock draws is sufficient to generate the covariance matrices used in this study. 
In Figure \ref{fig:covar_conv}, 500000 mock draws is represented by the vertical dashed grey line.

\section{Non-Gaussian distribution of the values of the auto-correlation function} \label{appendix:multi gaussian}

For our inference, we used the multi-variate Gaussian likelihood defined in equation \eqref{eq:gauss_like}. 
This functional form assumes that the distribution of mock draws of the auto-correlation function is Gaussian distributed about the mean for each bin.
In order to visually check this we will look at the distribution of mock draws from two bins of the auto-correlation function for two different models. 

\begin{figure}
	\includegraphics[width=\columnwidth]{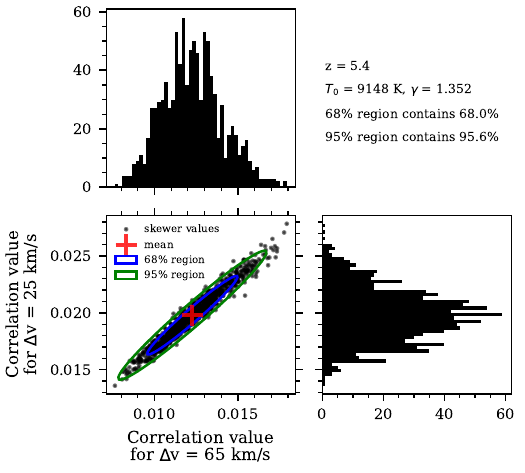}
    \caption{
        This figure shows the distribution 1000 mock draws from two bins of the auto-correlation function ($\Delta v = \SI{25.0}{\kilo\meter\per\second}$ and $\Delta v = \SI{65.0}{\kilo\meter\per\second}$) for one model ($z = 5.4$, $T_0 = 9148$ K, $\gamma = 1.352$, and $\langle F \rangle = 0.0801$).
        The top panel shows the distribution of only the $\Delta v = \SI{65.0}{\kilo\meter\per\second}$ bin while the right panel shows the distribution of only the $\Delta v = \SI{25.0}{\kilo\meter\per\second}$ bin. 
        The blue (green) circle represents the 68\% (95\%) ellipse calculated from the covariance matrix calculated for this model from equation \eqref{eq:covariance}. 
        The red plus shows the calculated mean. 
        Additionally the percent of mock draws that fall within each of these contours is written in the top right. 
        Both the 1D and 2D distributions seem relatively well described by a Gaussian distribution. 
        In the 2D plot, there are more points outside the 95\% contour to the top right than on any other side but it is not extreme. 
    }
    \label{fig:app_gaus_corner_z54}
\end{figure}

\begin{figure}
	\includegraphics[width=\columnwidth]{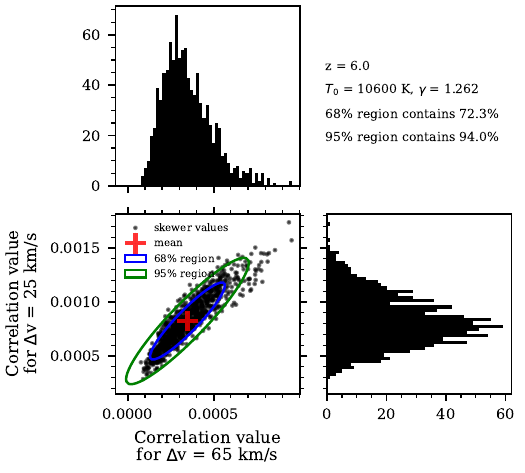}
    \caption{
        This figure shows the distribution 1000 mock draws from two bins of the auto-correlation function ($\Delta v = \SI{25.0}{\kilo\meter\per\second}$ and $\Delta v = \SI{65.0}{\kilo\meter\per\second}$) for one model ($z = 6$, $T_0 = 10600$ K, $\gamma = 1.262$, and $\langle F \rangle = 0.0089$).
        The top panel shows the distribution of only the $\Delta v = \SI{65.0}{\kilo\meter\per\second}$ bin while the right panel shows the distribution of only the $\Delta v = \SI{25.0}{\kilo\meter\per\second}$ bin. 
        The blue (green) circle represents the 68\% (95\%) ellipse calculated from the covariance matrix calculated for this model from equation \eqref{eq:covariance}. 
        The red plus shows the calculated mean. 
        Additionally the percent of mock draws that fall within each of these contours is written in the top right. 
        Both the 1D and 2D distributions are not well described by a Gaussian with 72.3\% of the mock draws falling within the 68\% contour and 94.\% of the mock draws falling within the 95\% contour.
    }
    \label{fig:app_gaus_corner_z6}
\end{figure}

Both Figures \ref{fig:app_gaus_corner_z54} and \ref{fig:app_gaus_corner_z6} show the distribution of 1000 mock data sets from the velocity bins of the auto-correlation function with $\Delta v = \SI{25.0}{\kilo\meter\per\second}$ and $\Delta v = \SI{65.0}{\kilo\meter\per\second}$.
The bottom left panels show the 2D distribution of the auto-correlation values from these bins. 
The blue (green) ellipses represents the theoretical 68\% (95\%) percentile contour calculated from the covariance matrix calculated for each model from equation \eqref{eq:covariance}. 
The red crosses shows the calculated mean. 
The top panels show the distribution of only the $v = \SI{65.0}{\kilo\meter\per\second}$ bins while the right panels show the distribution of only the $v = \SI{25.0}{\kilo\meter\per\second}$ bins. 

Figure \ref{fig:app_gaus_corner_z54} shows mock values of two bins of the auto-correlation function for the model at $z = 5.4$ with $T_0 = 9148$ K, $\gamma = 1.352$, and $\langle F \rangle = 0.0801$. 
Both the 1D and 2D distributions seem relatively well described by Gaussian distributions by eye though they do show some evidence of non-Gaussian tails to larger values. 
The number of points falling in each contour both fall within 1\% of the expected values. 
In the bottom left panel with the 2D distribution there are more mock values falling outside the 95\% contour to the top right (higher values) than in any other direction.
For this reason the distribution is not exactly Gaussian but a Gaussian visually appears as an acceptable approximation.

Figure \ref{fig:app_gaus_corner_z6} shows mock values of two bins of the auto-correlation function for the model at $z = 6$ with $T_0 = 10600$ K, $\gamma = 1.262$, and $\langle F \rangle = 0.0089$. 
In both the top and right panels, which show the distribution of values for one bin of the auto-correlation function, the distribution of mock draws is skewed with tails to the right. 
This is quantified by the percent of points in the two ellipses from the bottom left panel labeled in the top right with 72.3\% of the mock draws falling within the 68\% contour and 94.0\% of the mock draws falling within the 95\% contour.
The points outside of the contours are highly skewered towards the top right (higher values). 
It is only possible for the auto-correlation function to be negative due to noise, which generally averages to very small values approaching zero at the non-zero lags of the auto-correlation function. 
This can be seen in the black points and histogram do not go below 0, though the 95\% ellipse shown in green in the bottom left panel does go negative for $\Delta v = \SI{65}{\kilo\meter\per\second}$. 

Figures \ref{fig:app_gaus_corner_z54} and \ref{fig:app_gaus_corner_z6} show the changing distribution of the auto-correlation value with $z$, $T_0$, $\gamma$, and $\langle F \rangle$. 
There is a greater deviation from a multi-variate Gaussian distribution at higher $z$.
It is possible that adding additional sightlines will cause the auto-correlation function to better follow a multi-variate Gaussian distribution due to the central limit theorem, though investigating this in detail is beyond the scope of the paper. 
However, even with more sightlines $\langle F \rangle$ will be low at high-$z$ so we still expect the distribution to be skewed as the values mainly will not fall below 0. 
The incorrect assumption of the multi-variate Gaussian likelihood thus contributes to the failure of our method to pass an inference test as discussed in Section \ref{section: inf re-weight} for $z = 5.4$ and Appendix \ref{appendix:inference z6} for $z = 6$. 
For our final constraints, we calculated weights for our MCMC chains such that the resulting posteriors do pass our inference test, as discussed in Section \ref{section: inf re-weight}.
The whole method of assuming a multi-variate Gaussian then re-weighting the posteriors in non-optimal and future work using a more correct likelihood or likelihood-free inference will improve our results.

\section{Inference test at high redshift} \label{appendix:inference z6}

\begin{figure*}
    \centering
    \begin{subfigure}[t]{0.49\textwidth}
        \centering
        \includegraphics[width=\linewidth]{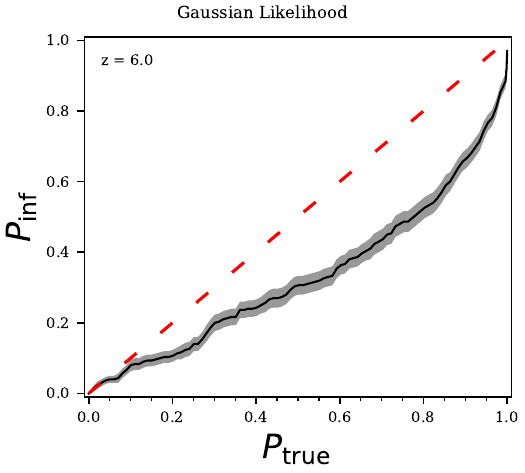} 
    \end{subfigure}
    \hfill
    \begin{subfigure}[t]{0.49\textwidth}
        \centering
        \includegraphics[width=\linewidth]{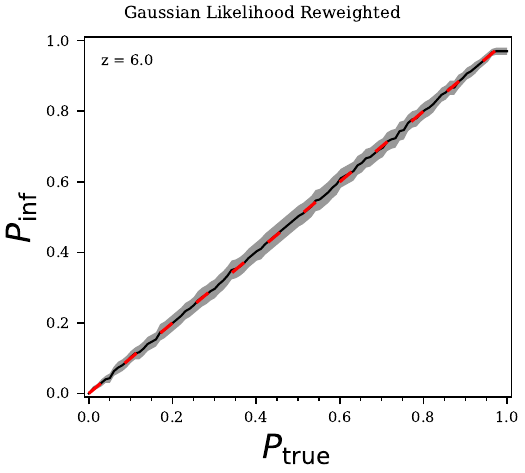} 
    \end{subfigure}
    \caption{
        The left panel of this figure shows the coverage resulting from the inference test from 300 models at $z = 6.$ drawn from our priors on $T_0$, $\gamma$, and $\langle F \rangle$.
        Here we see that the true parameters for the models fall above the 60th percentile in the MCMC chain $\sim 35\%$ of the time, for example. 
        The right panel of this figure shows the coverage resulting from the inference test with the use of one set of weights to re-weight the posteriors. 
        With these weights we are able to pass the inference test. 
    }
    \label{fig:inf_lines_z6}
\end{figure*}

Here we present the results of the inference test at $z = 6$. 
This calculation was done following the procedure described in Section \ref{section: inf re-weight}. 
Figure \ref{fig:inf_lines_z6} shows the results for $z = 6$ and can be compared to the $z = 5.4$ results in Figure \ref{fig:inf_lines}. 
The left panel here shows the initial coverage plot which deviates greatly from the expected $P_{\text{inf}} = P_{\text{true}}$ line, much more so than the $z = 5.4$. 
This likely comes from a greater deviation from the assumption of a multi-variate Gaussian likelihood as described in Appendix \ref{appendix:multi gaussian}. 
The $z = 6$ mock data show highly skewed distributions that are not well described by a Gaussian likelihood. 
The inference lines at other redshifts are available upon request.

\section{Gaussian data inference test} \label{appendix: gauss data inf}

\begin{figure*}
    \centering
    \begin{subfigure}[t]{0.49\textwidth}
        \centering
        \includegraphics[width=\linewidth]{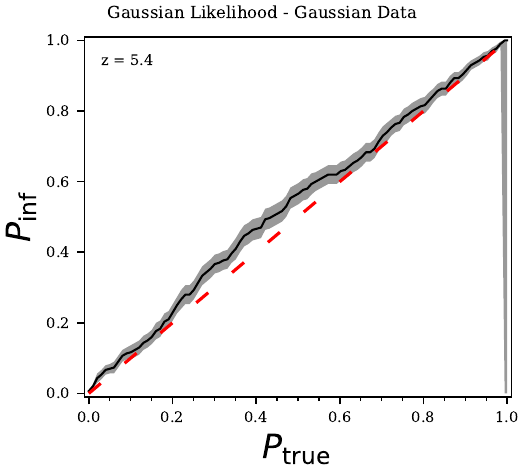} 
    \end{subfigure}
    \hfill
    \begin{subfigure}[t]{0.49\textwidth}
        \centering
        \includegraphics[width=\linewidth]{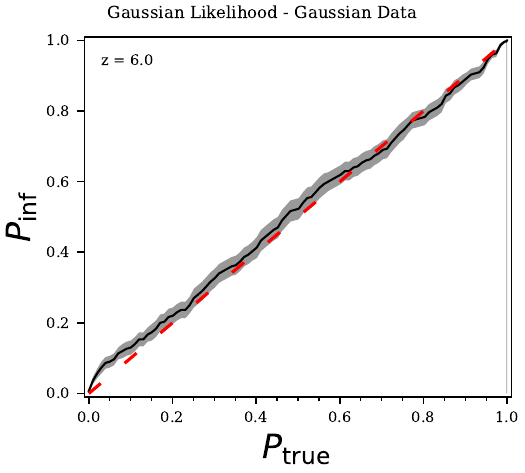} 
    \end{subfigure}
    \caption{
        Both panels of this figure shows the coverage plot resulting from the inference test from 300 data sets generated by randomly drawing points from the mean model and covariance matrix. 
        The the means and covariance matrices used come from $z = 5.4$ in the left panel and $z = 6.0$ in the right panel. 
        The true parameter values for both panels were drawn from our priors on $T_0$, $\gamma$, and $\langle F \rangle$.
        In both panels, the Gaussian mock data produced inference lines that fall on top of the 1-1 line within errors, as expected for the statistically correct posteriors. 
    }
    \label{fig:inf_lines_gauss}
\end{figure*}

As shown in Appendix \ref{appendix:multi gaussian}, the distribution of mock values of the auto-correlation function is not exactly Gaussian distributed. 
In order to confirm the failure of our mock data to pass an inference test (as discussed in Section \ref{section: inf re-weight} and Appendix \ref{appendix:inference z6}) comes from the use of a multi-variate Gaussian likelihood, we generate Gaussian distributed data and run inference tests. 
For one value of $T_0$, $\gamma$, and $\langle F \rangle$, we randomly generate a mock data set from a multi-variate Gaussian with the given mean model and covariance matrix calculated from our mock data as described in Section \ref{sec:autocorr}. 
We can then continue with the inference test as described in Section \ref{section: inf re-weight}. 
The results for this inference test for $z = 5.4$ and $z = 6.0$ are shown in Figure \ref{fig:inf_lines_gauss}. 
Here both redshifts inference lines fall along the 1-1 line that is expected for all probability contour, $P_{\text{true}}$, values. 
This behavior is also seen at the other redshifts. 
The fact that perfectly Gaussian data passes an inference test with the same likelihood, priors, and method as was used on mock data confirms that the failure of our mock data to pass an inference test is due to the non-Gaussian distribution of the mock data.

%%%%%%%%%%%%%%%%%%%%%%%%%%%%%%%%%%%%%%%%%%%%%%%%%%

% Don't change these lines
\bsp	% typesetting comment
\label{lastpage}
\end{document}